\documentclass[]{aa}

\usepackage{amsmath} 
\usepackage{amssymb}

\usepackage{txfonts}

\usepackage[utf8]{inputenc}
\usepackage{epsfig}
\usepackage{url}
\usepackage{color}
\definecolor{darkgreen}{rgb}{0,.6,0}
\definecolor{linkcol}{rgb}{.6,0,0}
\usepackage[colorlinks=true,citecolor=darkgreen]{hyperref}

\usepackage{natbib}
\bibpunct{(}{)}{;}{a}{}{,}
\usepackage{subfigure}

\usepackage{upgreek}  
\newcommand\unit[1]{\,{\rm #1}} 
\newcommand\inv[1][1]{\ensuremath{^{-#1}}}  

\newcommand\vmax[1][]{{\ensuremath{v#1_{\rm max}}}}
\def\tidalE{\ensuremath{E_{\rm t}}}

\def\totE{\ensuremath{E}}
\newcommand\mean[1]{\langle {#1}\rangle}

\def\surf{\ensuremath{I}}

\def\eg{{\it e.g.}}
\def\ie{{\it i.e.}}

\def\inner{{\it inner}}
\def\intermediate{{\it intermediate}}
\def\outer{{\it outer}}

\def\mass{{\cal M}} 
\def\msol{\ensuremath{\rm \mass_\odot}} 
\newcommand\solm\msol

\usepackage{dcolumn} \newcolumntype{d}[1]{D{.}{.}{#1}}
\usepackage{multirow}

\title{The Destruction of an Oort Cloud in a rich stellar cluster}
\author{T. Nordlander\inst1 \fnmsep \thanks{\emph{Current address:} Research School of Astronomy and Astrophysics, Australian National University, ACT 2611, Australia. \email{thomasn@mso.anu.edu.au}} \and H. Rickman\inst1 \fnmsep \inst 2 \and B. Gustafsson\inst1 \fnmsep \inst 3}
\institute{Division of Astronomy and Space Physics, Department of Physics and Astronomy, Uppsala University, Box 516, 75120 Uppsala, Sweden  \\ \email{Thomas.Nordlander@physics.uu.se}
	\and PAN Space Research Center, Bartycka 18A, 00716 Warszawa, Poland 
	\and Nordita, Roslagstullsbacken 23, 10691 Stockholm, Sweden
}

\abstract{It is possible that the formation of the Oort Cloud dates back to the earliest epochs of solar system history. At that time, the Sun was almost certainly a member of the stellar cluster, where it was born. Since the solar birth cluster is likely to have been massive ($10^3 - 10^4$\unit{\msol}), and therefore long-lived, an issue concerns the survival of such a primordial Oort Cloud.} 
{We have investigated this issue by simulating the orbital evolution of Oort Cloud comets for several hundred Myr, assuming the Sun to start its life as a typical member of such a massive cluster.} 
{We have devised a synthetic representation of the relevant dynamics, where the cluster potential is represented by a King model, and about 20 close encounters with individual cluster stars are selected and integrated based on the solar orbit and the cluster structure. Thousands of individual simulations are made, each including 3\,000 comets with orbits with three different initial semi-major axes.} 
{Practically the entire initial Oort Cloud is found to be lost for our choice of semi-major axes ($5\,000-20\,000$\unit{au}), independent of the cluster mass, although the chance of survival is better for the smaller cluster, since 
in a certain fraction of the simulations the Sun orbits at relatively safe distances from the dense cluster centre.
} 
{For the range of birth cluster sizes that we investigate, a primordial Oort Cloud will likely survive only as a small inner core with semi-major axes $\lesssim 3\,000$\unit{au}. Such a population of comets would be inert to orbital diffusion into an outer halo and subsequent injection into observable orbits. Some mechanism is therefore needed to accomplish this transfer, in case the Oort Cloud is primordial and the birth cluster did not have a low mass. From this point of view, our results lend some support to a delayed formation of the Oort Cloud, that occurred after the Sun had left its birth cluster.} 


\date{Received 23 December 2016 / Accepted 07 April 04 2017}

\begin{document}
\maketitle


\section{Introduction}

The formation of the Oort Cloud is one of the important issues when trying to understand the origin and evolution of the Solar System. This has been the case ever since this structure was first recognised \citep{Oort1950}, and resolving the issue still presents a very difficult task. It is natural to think of a ``primordial'' origin connected to the formation of the planets during the earliest stages of the Solar System more than 4.5\unit{Gyr} ago, as did Oort himself, and this has led to the classical picture \citep{Duncan1987,Dones2004} of comets as icy planetesimals scattered through the gravity of the growing giant planets into orbits extending far enough to sometimes be decoupled from the planetary system by external agents (Galactic tide and passing stars).

A different scenario for the Oort Cloud formation was recently investigated by \citet{brasser2013}. They explored one of the consequences of the Nice Model \citep[][and references therein]{Levison2011} for the long-term dynamical evolution of the giant planets. As a result of the rapid migration of Uranus and Neptune through the primordial trans-planetary disk into their current orbits, a scattered disk would be formed and, as an unavoidable by-product, also an Oort Cloud. This suggestion places the origin of the cloud at the time of the Late Heavy Bombardment (LHB) about 4\unit{Gyr} ago.

As long as this version of the Nice Model stands, there is strictly no need for the Oort Cloud to include any primordial component. However, even so, such a component is not ruled out. Moreover, the Nice Model may also accommodate a different scenario, where the planet migration happens very early. In this case, the Oort Cloud would definitely be primordial, and hence this option needs to be considered. An important issue then concerns the efficiency in the transfer of planetesimals into the Oort Cloud. There are two steps involved: first, the scattering of planetesimals into orbits that may be modified by external actions, and second, the decoupling that causes storage into the Oort Cloud.

It has been realised -- ever since the work of \citet{Gaidos1995} and \citet{Fernandez1997} -- that the Oort Cloud storage is strongly dependent on whether one treats the new-born Sun as an isolated star or a member of a dense stellar environment in a so-called birth cluster. The latter offers a more efficient way to decouple the objects by raising their perihelion distances due to the frequent occurrence of close and slow stellar encounters.

A numerical study of the formation of the Oort Cloud in a stellar cluster was performed by \citet{FernBrun2000}. In this work, the cluster was assumed to exist for a period of 100\unit{Myr} with a stepwise decreasing number density of stars from the assumed initial value down to zero. In addition, the tidal effect of the placental molecular cloud gas was included, typically only for the first 10\unit{Myr}. The scattering and decoupling of comets was simulated with the main result that an Oort Cloud inner core was formed quite rapidly with Jupiter and Saturn as the main scattering agents.

The discovery in 2003 of (90377) Sedna, whose perihelion distance of 76\unit{au} is well beyond the orbits of the giant planets and thus can only be explained by the influence of external actions,
spurred an interest in very dense stellar environments for the birth of the Solar System. 
Embedded clusters were recognised to be a common birth place for solar type stars \citep{Lada2003}. \citet{brasser2006} found a good match to the Sedna orbit for an inner Oort Cloud obtained in a model, where the Sun was born in such a cluster with a very high mass density. However, in a follow-up paper, \citet{brasser2007} investigated the effects of gas drag from the solar nebula on the planetesimal scattering and found what they referred to as size sorting. Only very large objects would evolve in the way described by \citet{brasser2006}, while orbits of comet-sized objects (radii $\sim 1$\unit{km}) would be circularized beyond the planetary orbits, opposing their scattering.

In a paper by \citet{KaibQuinn2008}, the formation of the Oort Cloud in a stellar cluster environment was again considered with similar assumptions for the cluster lifetime and number density of stars. However, the evolution thus computed during the first 100\unit{Myr} was supplemented by an additional 4.4\unit{Gyr} in a model of the galactic disk tide plus field star encounters. The effect of the resulting present Oort Cloud being dominated by a tight, inner core was confirmed as well as the possibility of obtaining a Sedna-type population due to the random effects of the closest stellar encounters.

An often cited model for the formation of a primordial Oort Cloud was proposed by \citet{levison2010}. This model assumes a birth cluster with few stars ($30<N<300$) and thus with a short lifetime. In this case, the stars are found to fly apart, when the gas component of the system is purged due to external influences. The Oort Cloud is formed by scattered disk comets from different stars of the cluster and gets enriched during the cluster break-up. However, the above-mentioned problem of bringing kilometre-sized objects into extended scattered disks was not addressed.

The current models of Oort Cloud formation may thus be summarized as, on the one hand, models for a primordial cloud, assumed to be formed in a dense but more or less short-lived stellar environment, and on the other hand, a delayed formation model associated with the LHB, where the Solar System is assumed to have left its birth cluster at an earlier stage. All these models assume a rather short dissolution lifetime for the birth cluster, which may or may not be true.

The size of the Sun's birth cluster was discussed in a review by \citet{adams2010}. His analysis of a range of constraints led to a broad probability distribution for the number $N$ of stars, peaking at $N \simeq 2\,500$. An important argument against too small values of $N$ was that it would be too unlikely for the birth cluster to produce a supernova with a progenitor mass of 25\unit{\msol} or more, as seems necessary to explain the amounts of short-lived radio isotopes that meteorite evidence show to have been present in the solar nebula \citep[\eg,][]{williams2007}. However, in itself this factor gives rather weak constraints -- only clusters with $N < 50$ would be excluded, since the random likelihood of the supernova would then be less than 5\%. The main factor opposing too large $N$ values was a too small chance for the regularity of the giant planet orbits to survive in the presence of the close stellar encounters then implied (\citealt{adams2001}; see also \citealt{malmberg2007}).

More recently, \citet{Gounelle2012} proposed a model for the origin of the short-lived radionuclides involving two generations of stars, formed in the same giant molecular cloud and preceding the formation of the Sun. This model implies a Solar birth environment rich in stars. Several thousand stars were estimated to have been born before the Sun, thus providing the source for $^{26}$Al and $^{60}$Fe traced in chondritic meteorites. The dynamical fate of the Sun was not addressed, but it seems possible that the Sun stayed gravitationally bound to the initial complex of stars and gas, thus becoming a member of a massive stellar cluster.

Thus, the Sun's birth cluster may have been rich in stars, containing thousands or more to begin with. Moreover, if we leave aside the specific constraints posed by solar system evidence and consider the statistics of observed embedded clusters, we find that the number of clusters formed today falls off with the number of member stars in such a way that about equal numbers of stars form in clusters with $10^2$, $10^3$ and $10^4$ members \citep{Lada2003}. While several papers have dealt with Oort Cloud formation in a cluster with $\sim10^2$ members, which we shall call a low-mass (LM) cluster, the other two classes of clusters -- intermediate-mass (IM) and high-mass (HM) clusters -- have not yet been considered. The cases treated by \citet{FernBrun2000} and \citet{KaibQuinn2008}, where the whole cluster dissolves within 100\unit{Myr}, would fall into our LM category.

Dynamical models of stellar cluster evolution show that clusters with more than 1\,000 initial members typically survive for several hundred Myr or more \citep{Lamers2008}. Thus, their lifetimes may even exceed the interval from the formation of the earliest solar nebula condensates (meteoritic calcium-aluminium rich inclusions or CAI) until the triggering of the LHB \citep{morbidelli2012a}. This calls for a reevaluation of the Oort Cloud formation models in the framework of such a long-lived birth cluster. The large number of stars may help in the formation of a primordial cloud, but it is also a threat to the stability of the cloud due to the possibly disruptive effect of subsequent encounters. In the present paper we consider the fate of a primordial Oort Cloud in a dense stellar environment that lasts at least until the LHB. The main question is if such a cloud would survive or not.

We use two assumptions for the birth cluster, considering two values for the initial number of stars ($N_0$). For the IM cluster we take $N_0 = 2\,000$, and for the HM cluster we choose Messier~67 as a template. In this case, $N_0 = 36\,000$ whereof half the systems are binary \citep{hurley2005}. This choice is arbitrary, and we do not mean to suggest M67 to be the Sun's birth cluster -- for discussions of this issue, see \citet{pichardo2012} and \citet{gustafsson2016}. It is, however, a convenient case for the upper end of the mass range due to the availability of detailed evolutionary modelling based on a very good observational record.

We have devised a modelling technique that allows to trace the motions of thousands of test objects representing Oort Cloud comets on heliocentric orbits for a time span of several hundred Myr, in the presence of a static cluster potential plus a selection of randomly created stellar encounters. These are meant to include some of those that impart the largest impulses to the Sun. The CPU time consumption is moderate enough to allow running thousands of such integrations for each model of the birth cluster and thus obtaining results that are robust against statistical uncertainties.

In Sect.~\ref{sec:methods} we first present our cluster model and describe its two versions (IM and HM) in some detail. We then describe our treatment of stellar encounters and our derivation of the encounter frequency and velocity distribution. Next, we present our simulation set-up. The results are given in Sect.~\ref{sec:results}. Notably, we find that the survival probability of primordial Oort Cloud comets is extremely low regardless of the size of the birth cluster, within the cluster mass range explored. Conclusions drawn from this result and a discussion are given in Sect.~\ref{sec:discus}. In two appendices we describe the calculation of the cluster model structure and the implementation of the stellar encounters, respectively.



\section{Methods} \label{sec:methods}
We integrate the orbit of the Sun together with thousands of test particles representing Oort Cloud comets, in a static cluster potential over the course of 400\,Myr. The comets are introduced in random heliocentric orbits with semi-major axes of $a_{\rm o} = 5\,000$, $10\,000$, and $20\,000\unit{au}$.
The cluster potential is computed for an intermediate-mass (IM) cluster based on a template with an initial number of $N_0 = 2000$ stars, and a high-mass (HM) cluster whose template initially had $N_0 = 36\,000$ stars, as described in Sect.~\ref{sec:cluster}. 
We represent the HM cluster by one static model while we use a sequence of static models for the IM cluster, as this evolves rapidly with time and in fact nearly dissolves during the simulated time period.

During each simulation, we select roughly 20 stellar encounters to occur at random times, by means of an impact approximation described in Sect.~\ref{sec:encounterflux}. We add these interloping stars one at a time to the orbit integrations as described in Sect.~\ref{sec:encounterselect}. The gravitational influence on the comets is thus that of the smooth cluster potential, the Sun, and at times, one additional star.
The numerical setup and the selection of comets is further described in Sect.~\ref{sec:simulations}.

\subsection{General Properties of the cluster model}
\label{sec:cluster}
\label{sec:cl-prop}

We impose the following constraints on our cluster model. 
\begin{itemize}
\item{The stellar distribution function corresponds to a spatially isotropic, relaxed state.}
\item{It does not directly account for mass segregation and considers only single stars.}
\item{It is stepwise or fully constant with time during the interval we simulate.}
\end{itemize}
These constraints are to some extent mutually incompatible, since mass 
segregation and binary formation are necessary consequences of the same dynamics that causes
relaxation. In addition, for the relatively young system that we simulate, relaxation should still be 
ongoing. Thus, the system has to evolve with time in contrast to our third constraint.
The reason why we stick to the above concepts is that they allow us to develop a synthetic 
model of stellar encounters that greatly facilitates the Oort Cloud simulations and avoids the use 
of time-consuming $N$-body simulations.

Clearly, a most useful template in order to achieve a cluster model with the above properties 
is the \citeauthor{king1966} model \citep{king1966}. We apply this concept according to a 
prescription that we describe in Appendix~\ref{sec:stellarstructure}.

To fit our model parameters, we start from an initial guess (see Appendix~\ref{sec:stellarstructure}), and improve these values iteratively, until they yield the desired solution. This means that the cluster has the required mass, and the density function drops to zero at a distance $R_{\rm lim}$ less than the tidal radius. Other fitted parameters include the half-mass radius $r_{\rm h}$. Finally, we may also compute the sky projected surface density $\surf(R)$ as prescribed by \citet[eqns. 23--27]{king1966}. 
Using this quantity, we calculate the core radius $R_{\rm c}$ from its definition $\surf(R_{\rm c}) = \frac 1 2 \surf(0)$. In our case, the surface density refers to mass rather than brightness. This core radius may also be used as a fitting parameter.

The last property to be defined is the age of the cluster. We want to simulate the cluster effects on the Oort Cloud for a time interval, extending from the formation of a primordial Cloud until the LHB. 
However, in this work we do not consider the effects of the cluster on the very formation of the cloud. In reality, the formation process is expected to extend over several hundred Myr \citep{KaibQuinn2008,brasser2013}, but we replace this by a step function: we consider a first time interval of 100\unit{Myr} starting at the formation of the solar system ($t=0$), during which there is no Oort Cloud at all. Then, at $t=100$\unit{Myr}, we introduce the comets in their above-described, initial orbits. The cluster age hence extends from 100\unit{Myr} to 500\unit{Myr}, when we assume the LHB to occur.

\subsubsection{The High-Mass Cluster}\label{sec:cl-him}

As indicated above, the structural parameters of this cluster are based on the properties of a young M67 as simulated by \citet{hurley2005} and available in their online tabulations\footnote{Data available (in March 2017) at \url{http://astronomy.swin.edu.au/~jhurley/nbody/archive.html}}. These authors performed a set of $N$-body simulations, describing the complete evolution of M67 after formation of all the stars and escape of any residual gas.
The initial parameters of their evolutionary model -- the galactocentric distance and orbit, total mass, and binary populations -- were estimated from the current luminous mass (\ie, the total mass of nuclear-burning stars, estimated at $\mass_{\rm L} \sim 1\,000 \unit{\msol}$), age (approximately 4.0\unit{Gyr}), and binary and blue straggler populations of the cluster. As these populations are sensitive to the dynamical evolution of the cluster, the model is reasonably well constrained.

\begin{figure}
\includegraphics[width=\hsize]{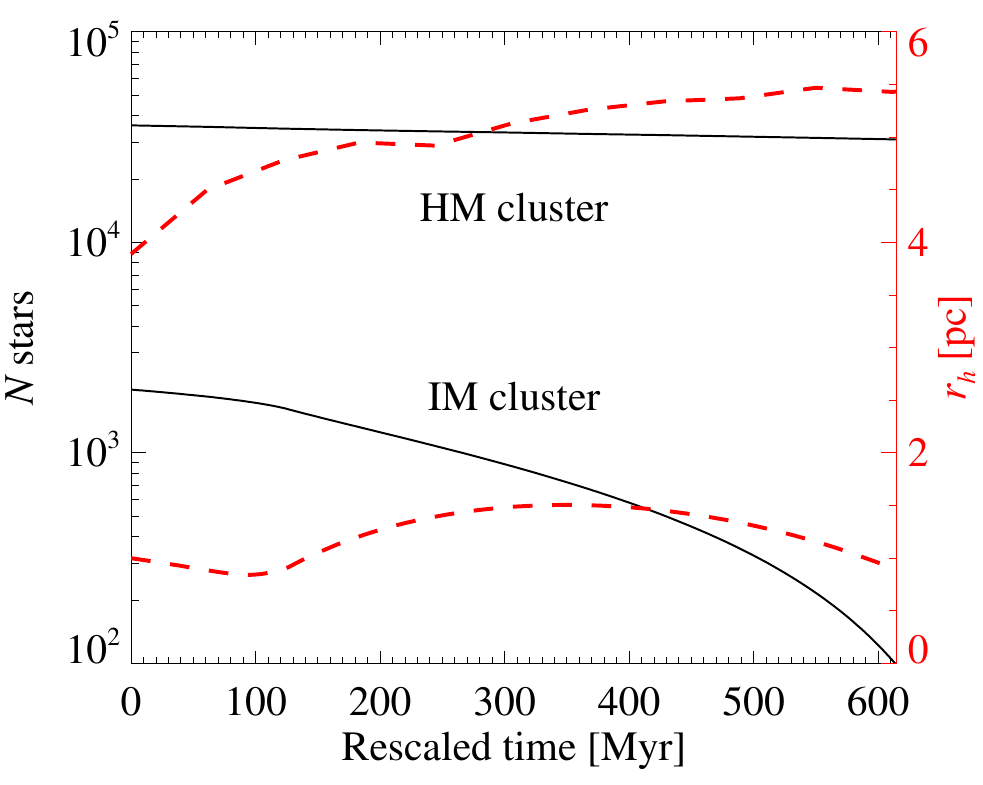}
\caption{The number of stars (solid, black lines) in our high-mass and intermediate-mass cluster models, and their half-mass radii (dashed, red lines), versus time from cluster formation. A logarithmic scale is used for the number of stars and a linear scale for the radius. The time has been rescaled for the IM cluster to take the expected influence of GMCs into consideration (see text).}
\label{fig:clmodels}
\end{figure}

In Fig.~\ref{fig:clmodels} we present the evolution of both our template clusters. The HM cluster, based on data from \citet{hurley2005}, is seen to decrease slowly in the number of stars during the considered time interval (from 100 to 500\unit{Myr}). We illustrate the total number of stars, which often occur as binary components in the real cluster M67 -- a discussion of binarity is given in Sect.~\ref{sec:cl-stars}. The inner part of the cluster undergoes a slight expansion with the half-mass radius increasing from an initial value of 4.0\unit{pc} to 5.5\unit{pc}.

The initial distributions of mass and energy used by \citet{hurley2005} were based upon a \citeauthor{plummer1911} model \citep{plummer1911}. On the half-mass relaxation timescale $T_{\rm rh,0} \approx 290$\unit{Myr}, this distribution evolves into something resembling a \citeauthor{king1966} profile. 
The relaxation causes the increase in the half-mass radius, which reflects the evolution of the density profile.

We choose a static setup for our model cluster representing the mean state of M67 during the 
age span 100--500\unit{Myr} in the simulations of \citet{hurley2005}. 
At this early stage in the cluster's history, the state can be described as semi-relaxed, so that the global
structure is neither that of the initial \citeauthor{plummer1911} distribution nor that of the eventual
\citeauthor{king1966} model.
However, the relaxation proceeds more rapidly in the inner parts of the cluster, where the crossing time is smaller. 
Therefore, the structure in this part of the cluster is less sensitive to the initial conditions of their model.
By choosing a \citeauthor{king1966} model, we are thus able to reproduce the inner parts of the simulated cluster reasonably well, while the outer parts are less well described.

The parameters to which we fit our \citeauthor{king1966} model are the
mass $\mass_{\rm cl}$, half-mass radius $r_{\rm h}$, core radius $R_{\rm c}$, limiting radius $R_{\rm lim}$, and the mean density within the half-mass and core radii, $\mean {\rho_{\rm h}}$ and $\mean {\rho_{\rm c}}$. 
The input parameters and resulting properties of the computed model are given in Table~\ref{tbl:modelparams}.

\subsubsection{The Intermediate-Mass Cluster}\label{sec:cl-lom}

While N-body simulations of the HM cluster indicated a rather slow evolution during the first 2--3\,Gyr, the IM cluster has a short relaxation time and thus rapidly evolves not only in total mass but also in terms of structure.
The half-mass relaxation time scales with the cluster mass and half-mass radius like $t_\text{rh} \propto \mass_\text{cl}^{1/2} r_\text{h}^{3/2}$ \citep{spitzer_random_1971}. 
For a population of $N_0 = 2\,000$ stars with $r_\text{h} = 1\unit{pc}$, we find $t_\text{rh} \approx 20 \unit{Myr}$, indicating that such a cluster should relax very rapidly.

In Fig.~\ref{fig:clmodels} we illustrate the simulated evolution of an IM cluster, computed using the \textsc{emacss} code \citep{alexander_prescription_2012,gieles_prescription_2014,alexander_prescription_2014}, with its lifetime rescaled according to the expected influence of encounters with GMCs \citep{gieles_star_2006}. 
We find that the relaxation time varies between 20 and 40\,Myr during the first 500\,Myr of the simulation.
The parabolic evolution of the half-mass radius after the initial relaxation is due to the combined effect of mass loss from the outer regions, and the collapse and bounce of the core, where energy transfer from the core inflates the outer regions of the cluster.

\begin{figure}
\includegraphics[width=\hsize]{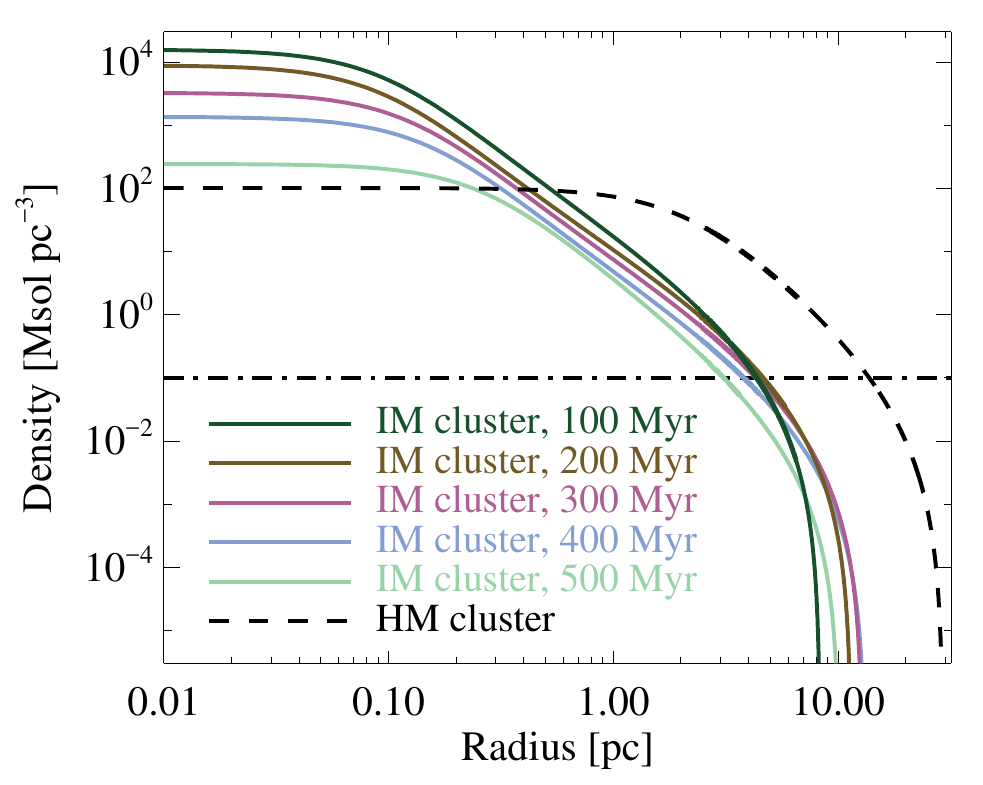}
\caption{Density distribution in the template clusters. 
The dashed black curve indicates the distribution of the high-mass cluster, while the solid coloured curves represent time steps in the intermediate-mass model, 100 Myr apart, with the average time of each step shown in the legend. The central mass density (left end of the plot) decreases monotonically with age of the model.
The horizontal dash-dotted line indicates the Galactic mid-plane density for comparison.
}
\label{fig:cluster_structure}
\end{figure}

Due to the rapidly evolving nature of this cluster, we represent it by a sequence of static models, each corresponding to the average properties during a span of 25\,Myr. The input parameters and resulting properties of the models are listed in Table~\ref{tbl:modelparams}.
The density distributions of the HM cluster model as well as every fourth model representing the IM cluster, are shown in Fig.~\ref{fig:cluster_structure}.
Both structures exhibit an inner plateau out to the core radius (1.9\,pc in the HM cluster; 0.05--0.1\,pc in the IM cluster), whereafter the density roughly follows a power-law, decreasing to values lower than that of the local Galactic disk.

\subsubsection{Cluster Stars}\label{sec:cl-stars}

In addition to the structure of the cluster, we need to describe its stellar content.
To translate a distribution of mass into the corresponding distribution of individual stars, 
we adopt a stellar IMF from the generating function of \citet{kroupa1993}, 
\begin{equation}
  \mass(\xi) = \mass_0 + \frac{0.19 \xi^{1.55} + 0.050 \xi^{0.6}}{(1-\xi)^{0.58}}
\end{equation}
where $\xi \in [0,1]$ is a random number from a uniform distribution, and $\mass_0$ the lower mass limit. 
We set $\mass_0 = 0.1 \unit{\msol}$ to account for the mass segregation-driven preferential loss of the lowest-mass objects from the cluster. 

We evolve the stellar population to an age of 300\unit{Myr}, representing the mean age of the cluster, 
using the rapid stellar evolution code SSE \citep{Hurley2000}. 
The resulting mean stellar mass is $\mean{\mass} \approx 0.41 \unit{\msol}$, with an effective upper mass limit near $3.5 \unit{\msol}$ (turnoff mass $\sim 3.3\unit{\msol}$). For completeness, we retain stellar remnants in the form of white dwarfs in the mass distribution, but remove neutron stars and stellar mass black holes as these are expected to usually be given kicks at formation much greater than the cluster escape velocity \citep[see, \eg,][]{pfahl2002}.

We do not directly invoke mass segregation in our cluster model. Thus, the probability distribution of stellar 
mass is the same at any distance from the centre. In this sense, our model fails to include a 
real phenomenon, which would be expected to occur in either cluster a few hundred Myr after its birth: a systematic 
tendency for the high-mass stars to concentrate in the cluster core and, hence, to be 
underrepresented in other parts of the cluster.

We also neglect the existence of binaries, even though these are common in M67 \citep{richer1998} and also in low-mass clusters \citep{giersz_statistics_1997}, since they would complicate the treatment of close encounters. If hard binaries were included, these 
would effectively increase the mean mass of the Oort Cloud perturbers while for a given cluster mass decreasing the overall number density of those perturbers. Note that \citet{fouchard2011} found the long term dynamics of the Oort Cloud to be influenced by massive stars in the Galactic disk to a much larger extent than their low encounter frequency would suggest -- the reason being their higher chance of producing global perturbing effects on the cloud. Hence, we may be underestimating the stellar encounter effects, but this is in line with our strategy to seek a conservative estimate.

\subsection{Stellar encounter flux} 
\label{sec:encounters}
\label{sec:encounterflux}

\begin{figure*}
\centerline{
\includegraphics[height=7.5cm]{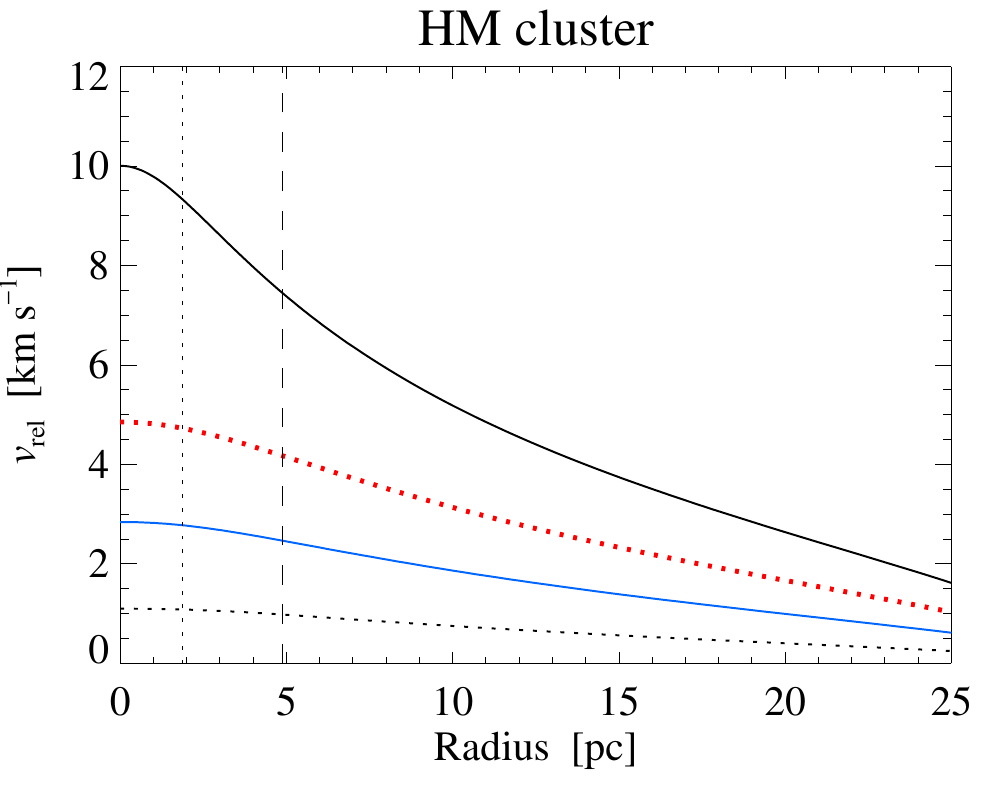}
\includegraphics[height=7.5cm,clip,trim=2em 0 0 0]{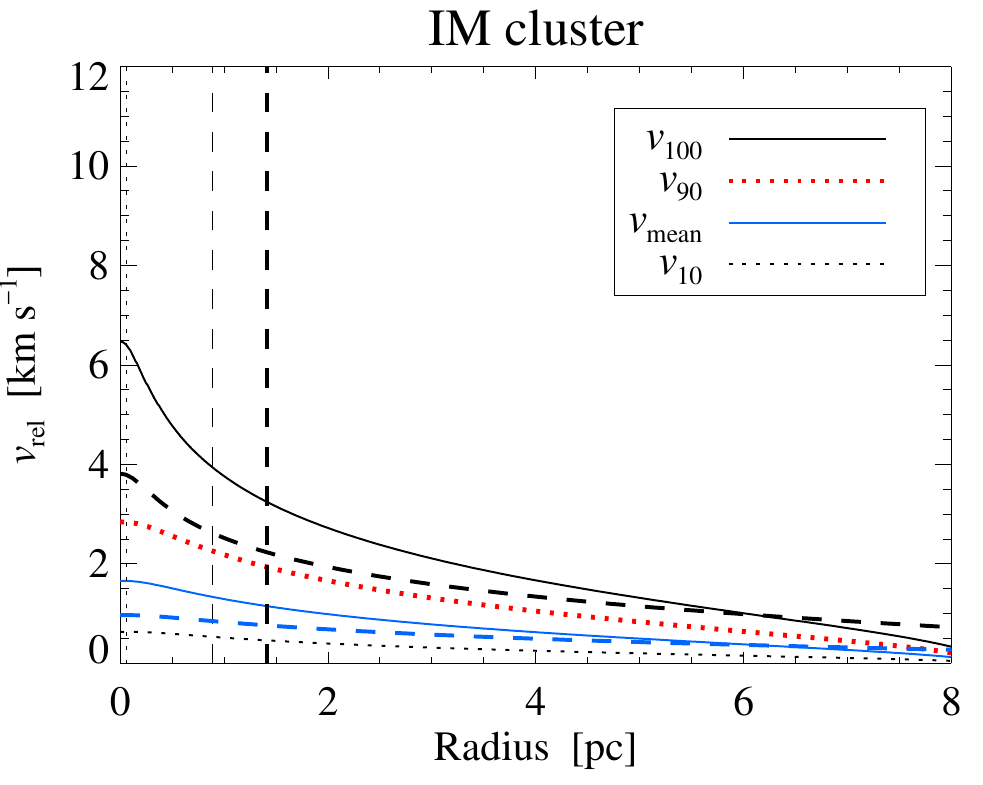}
}
\caption{Distribution of relative velocities in our encounter model for the stellar cluster. The five curves in the left panel represent, from top to bottom, the maximum allowed relative velocity (solid, black), its 90th percentile (dotted, red), mean (solid, blue) and 10th percentile values (dotted, black).
Shown as vertical (black) lines are the core (dotted) and half-mass (dashed) radii.
\textit{Left panel:} Relative velocity distributions for the high-mass cluster.
\textit{Right panel:} Relative velocity distributions for the intermediate-mass cluster at two time steps. The first step represents the time interval 100--125\,Myr. The second time step represents the time interval 300--325\,Myr, and is shown only by a thick dashed black line representing the maximum relative velocity, a thick dashed blue line representing the mean, and a vertical thick dashed line representing the half-mass radius at this time. The core radius is not shown at this time, but is available for all time steps in Table~\ref{tbl:modelparams}.
}
\label{fig:encounter_velocities}
\end{figure*}

Since we do not trace the motions of individual cluster stars, we generate encounters synthetically using a statistical encounter flux derived from the cluster model.

To derive this encounter flux, we first need the distribution of relative velocities of the stars.
At a given distance from the cluster centre, we calculate the cumulative distribution of kinetic energies from equation (\ref{eq:densityfunction}) and produce a generating function.
To represent individual encounters, we draw independently two random values from this generating function for the kinetic energy (per unit mass) of two encountering stars.

By assuming a flat distribution of angular momenta (as inherent in a King model), absorbing at most all of the available kinetic energy, we determine the tangential and radial velocity components of each star. Each radial velocity can be positive or negative with equal probabilities.
We assume an isotropic encounter distribution, wherein the motions of the two stars are uncorrelated.
The distribution of these encounter velocities is illustrated in Fig.~\ref{fig:encounter_velocities}, including the arithmetic mean $\langle v_{\rm rel}\rangle$ which in the figure is denoted $v_{\rm mean}$ and shown by blue curves. Note that when we use this to find the Sun's encounter flux, we do not account for the constraints imposed by the Sun's particular orbit. In our approximation, the Sun is treated as an average star at any particular distance from the cluster centre.

\begin{figure}
\includegraphics[width=\hsize]{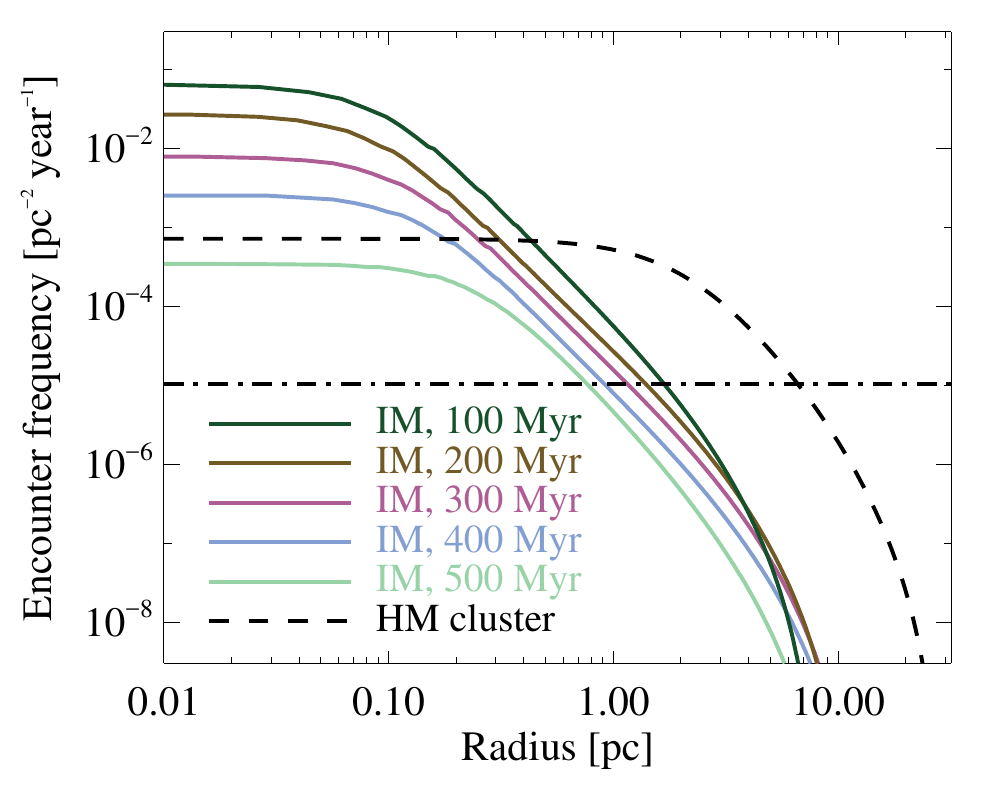}
\caption{Encounter frequency versus distance from the cluster centre, shown for the high-mass cluster (dashed line) and selected timesteps for the low-mass cluster (coloured solid lines). 
The encounter flux at the centre of the cluster decreases with increasing age for the intermediate-mass cluster -- colours are the same as in Fig.~\ref{fig:cluster_structure}.
The horizontal dash-dotted line illustrates for comparison the current encounter frequency of the Sun with Galactic field stars \citep{rickman2008}.
}
\label{fig:encounter_frequency}
\end{figure}

Using the mean relative velocities, we show in Fig.~\ref{fig:encounter_frequency}
the encounter frequency versus distance from the cluster centre: $f(r) = n \langle v_{\rm rel} \rangle$. 
For an assumed impact parameter $b$, this encounter frequency produces an encounter rate $\Gamma \equiv \sigma n \langle v_{\rm rel} \rangle$, where $\sigma \equiv \pi b^2$ is the impact cross-section. 
The expected number of encounters occurring over time $\tau$ with a constant value of the encounter rate is thus simply $\int_0^\tau \Gamma dt = \tau \Gamma$.
Taking $\Gamma = \tau^{-1}$ for a given time $\tau$, we may thus estimate the expected minimum impact parameter $b_{\rm min}$ of the encounters experienced during a timespan of length $\tau$. Using $\tau = 400\unit{Myr}$, 
in the HM cluster, we find a value of $b_{\rm min} \approx 210\unit{au}$ at the very centre, while near the half-mass radius we find $b_{\rm min} \approx 1200\unit{au}$.

For the IM cluster, we integrate $\int_0^\tau \Gamma(t) dt$, taking into account the time dependence of the encounter flux, by adding the contributions from the different 25\unit{Myr} intervals. At the cluster centre, we find $b_{\rm min} \simeq 35$\unit{au} for $\tau = 400$\unit{Myr}. Clearly, and not surprisingly, the Oort Cloud and even the planetary system would not survive such an extended stay in such a dense environment. If we take into account that the Sun is found to leave this kind of cluster before 100\unit{Myr} as a median (see Sect.~\ref{sec:solar_orbit}), for this value of $\tau$ we still have $b_{\rm min} \simeq 40$\unit{au}, which is very destructive. However, the solar orbit makes the Sun spend only a minimal amount of time in the immediate vicinity of the centre, if any at all. For a more realistic estimate, we also perform the estimate for a distance of 1.1\unit{pc} from the centre, roughly representing an average half-mass radius. In this case, the two values of $b_{\rm min}$ increase to approximately $1\,450$\unit{au} and $2\,000$\unit{au}, respectively. These are larger than the one found for the HM cluster.

The expected minimum impact parameter resembles the size of the Oort cloud, $b_{\rm min} \approx 10\,000\unit{au}$ at $r \approx 14\unit{pc}$ in the HM cluster, or at $r \approx 3.5\unit{pc}$ in the IM cluster. 
Note, however, that at these distances in either cluster, the cluster mass density is in fact comparable to that in the Galactic disk \citep[$\sim 0.1\unit{\msol\,pc\inv[3]}$,][]{holmberg2000}. 
In the disk, the mean encounter velocity is greater by an order of magnitude, which increases the flux of stellar encounters by the same amount to $10^{-5} \unit{pc\inv[2]\,Myr\inv}$, but also reduces the efficiency of momentum transfer. The influence of field star passages on the Oort Cloud has been investigated elsewhere \citep{rickman2008}, and the erosion of the Oort Cloud was then found not to be dramatic over intervals like the one considered here. Thus we neglect these passages as well as their associated Galactic disk tide effect.

\subsection{Stellar encounter selection} \label{sec:encounterselect}

We simulate stellar encounters by introducing a star at a distance $d_\text{start}$ from the Sun, and integrate the orbits of the two stars under their mutual gravitational influence in the cluster gravitational potential, until the mutual distance is $d_\text{end}$. 
In the simulation of the HM cluster, we found that setting $d_\text{start} = d_\text{end} = 1\unit{pc}$ results in a typical encounter duration less than 1\unit{Myr}, with a tail in the distribution of encounter durations extending toward 16 and 34\unit{Myr} representing fewer than five and one percent of encounters, respectively. Such extended encounters are similar in duration to the period of a Kepler orbit around the cluster centre near the half-mass radius in this cluster, 
$P \sim r_{\rm h}^{3/2} \mass_{\rm cl}^{-1/2} \sim 9\unit{Myr}$. 

In the simulation of the IM cluster, a distance $d_\text{start}$ of 1\unit{pc} would be inappropriate as this is in fact similar to the half-mass radius of the cluster. Thus, most of the interaction of the two stars occurs at distances much larger than those between the Sun and many other cluster stars, which makes the calculation somewhat irrelevant. In addition, as we shall explain below, we use a filter when selecting encounters to avoid overlaps of consecutive events. With the smaller relative velocities of the stars in the IM cluster, this would lead to the blocking of too many low-velocity encounters due to their large durations. 
To reduce this bias, we set a distance $d_\text{start} = 0.5\unit{pc}$ where the encountered star is introduced, 
and a slightly smaller distance $d_\text{end} = 0.45\unit{pc}$ where it is removed. 
This results in a similar distribution of encounter durations as in the case of the HM cluster.

We have checked that the particular choice of distances does not significantly influence the results by repeating our IM cluster calculations with $d_\text{start} = 0.25\unit{pc}$ and $d_\text{end} = 0.22\unit{pc}$, with no significant effect on either the survivability of comets or the evolution of the solar orbit. Similar tests on the HM cluster indicate that neither qualitative nor quantitative results depend strongly on the precise choice of the interaction distance.

The typical encounter in either simulation does not overlap subsequent pericentre passages, which is important as our setup does not allow for simultaneous encounters. We enforce this by a veto blocking the selection process while an encounter is ongoing. 
To avoid having the encounter scheme block too many subsequent encounters (typically occurring near pericentre), we aim for producing one encounter per 20\unit{Myr}, \ie, $\Gamma = (20\unit{Myr})^{-1}$. For a simulation duration of $\tau = 400\unit{Myr}$, this corresponds to typically 20 encounters per simulation, where the veto typically blocks one expected encounter per simulation. 

The number of stellar encounters expected to exert significant influence on the outer Oort Cloud is, however, larger than just 20 per 400\unit{Myr}, and we aim to simulate those encounters that are most important. Rather than using the distance as criterion, we adopt a strength parameter $S = \mass / (v_{\rm rel} b)$, approximating the impulse transferred to the Sun by a stellar encounter in the classical impulse approximation \citep[see][]{rickman1976,fouchard2011}. 
All three defining parameters must then be known for $S$ to be computed, which in turn requires the encounter geometry to be determined. 
As detailed in Appendix~\ref{sec:stellarencounters}, we pick the encounter parameters at random and generate a list of encounters likely to occur during a time interval of a given length, and compute $S$ for each. Finally, we select the encounter with the largest value of $S$.

When the final selection is made, the integration proceeds until the encounter is finished, as detailed in Sect.~\ref{sec:setup}. Then a new time interval of the given length is considered, using the modified solar orbit, and the next stellar encounter is selected. On the average, the interval between consecutive encounters will be close to 20\unit{Myr}, and hence a total of about 20 encounters will be treated during the entire 400\unit{Myr} simulation.

\subsection{Numerical Simulations} 
\label{sec:simulations}
\label{sec:setup}

We use a hybrid numerical setup. We combine the tidal effects of the cluster as a whole, represented by the smooth gravitational potential computed according to Sect.~\ref{sec:cluster}, with the influence of individual stars during orbit-integrated close encounters as described in Sect.~\ref{sec:encounters}. Comet motions are thus integrated under the gravitational influence of the cluster, the Sun, and an interloping star when applicable.
As mentioned above, we neglect the influence of Galactic tides and interloping Galactic field stars.

The Oort Cloud is represented by a sample of 3\,000 comets. This consists of three ensembles of 1\,000 comets each, representing different parts of an initial Oort Cloud, with original semi-major axes of
$a_{\rm o} = 5\,000$, $10\,000$, and $20\,000\unit{au}$. We shall here refer to the respective (initial) ensembles as the (initially) \inner, \intermediate\ and \outer\ comet cloud -- not to be confused with the nomenclature of actual present-day Oort Cloud populations.
The other orbital elements are drawn randomly from identical distributions for all three ensembles.
These distributions are uniform for all but the eccentricity, which is distributed as $f(e) \propto e$ in the range $e \in [0,1]$ to represent a thermalized state. Since our model does not include the planets, the initial cloud is modelled without any loss cone \citep{Oort1950,Hills1981}, thus allowing $e \to 1$. 
Orbital inclinations are given an isotropic distribution, uniform in $\cos i \in [-1, 1]$.

\begin{figure*}
\centerline{
	\includegraphics[width=.5\hsize]{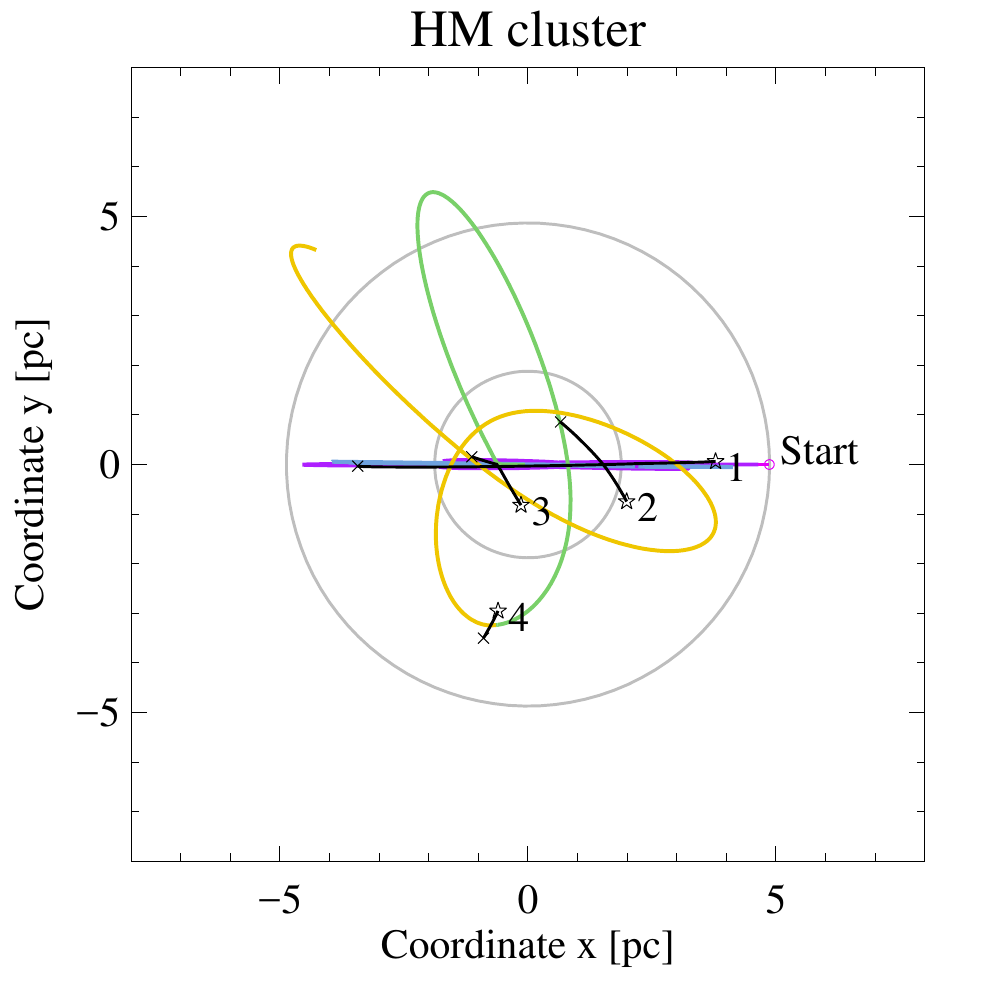}
	\includegraphics[width=.5\hsize]{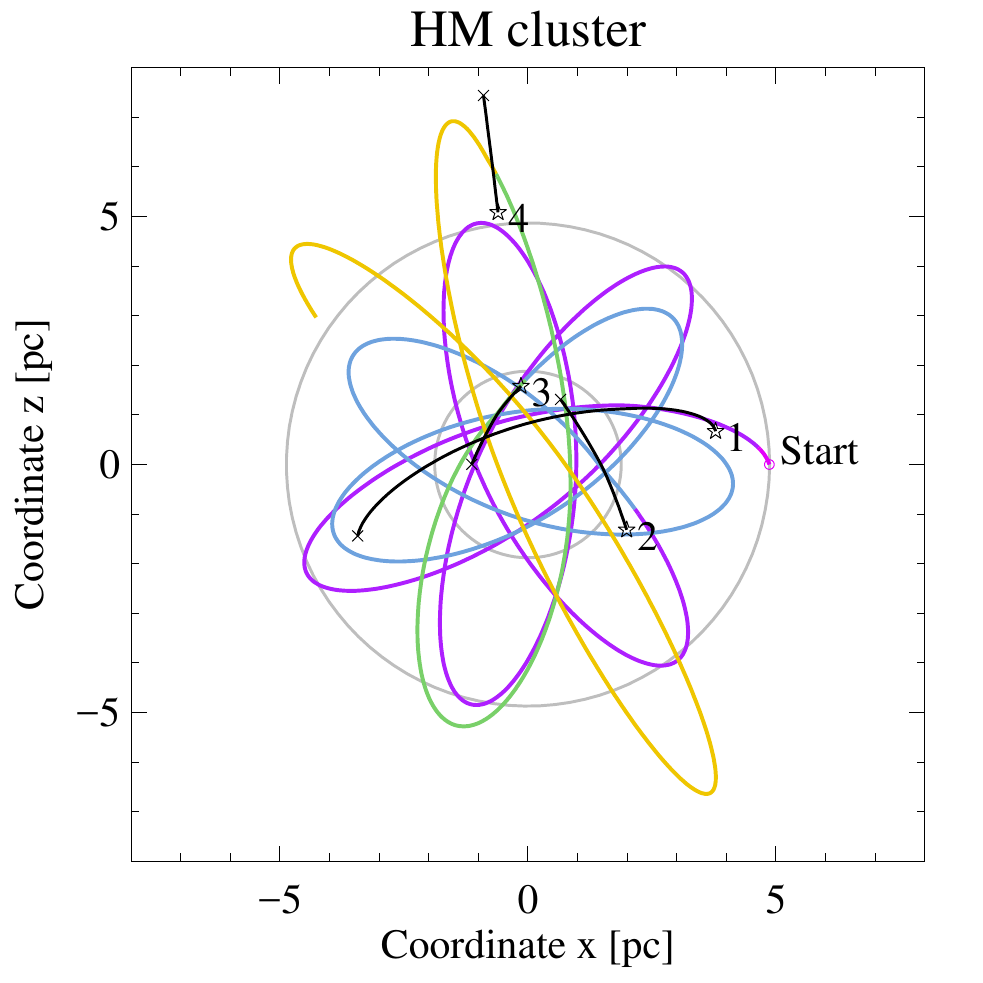}
}
\caption{Solar orbit during the first 80\,Myr in a simulation of the high-mass cluster, with grey circles indicating the core and half-mass radii. The solar orbit is illustrated by a line that switches colour (from violet to blue, green and yellow) every time a star is encountered. The starting points of the interloping stars are numbered and indicated by a star symbol, while end points are indicated by $x$. The starting point of the solar orbit is indicated by a circle of radius 20\,000\,au.
The solar orbit is by construction initially confined to the x-z plane, and departs significantly from that plane only after a close interaction with star number 3.
}
\label{fig:solarorbit}
\end{figure*}

The simulations are realised in the barycentric frame of the cluster, with the Sun initially positioned at the half-mass radius in the case of the HM cluster. In the IM cluster, we instead placed the Sun at a random distance from the cluster centre, distributed according to the cluster density profile. 
The initial velocity (and orbit) of the Sun are determined randomly from the energy and angular momentum distributions, and the time and configuration of the first (or next) encounter are determined as was described in Sect.~\ref{sec:encounterselect}.

The orbit of the Sun is integrated together with the comets until the initialization of the following encounter, where the star is introduced at a distance of 1\unit{pc} or 0.5\unit{pc} from the Sun for the HM or IM cluster, respectively. 
The integration then includes the mutual gravitational influence of the Sun and the star on each other, and on the comets, until the distance between the stars is again 1\unit{pc} or 0.45\unit{pc} for the HM or IM cluster, respectively. All the objects are also subject to the smooth gravity field of the cluster.
An example of the solar orbit during the first 80\,Myr of one of the simulations of the HM cluster is given in Fig.~\ref{fig:solarorbit}.

Comets which move beyond a distance of 1\unit{pc} from both the Sun and the star are logged and then discarded from the computations. In addition, following the results of \citet{fouchard2013}, we model a planetary loss cone such that comets reaching within 5\unit{au} of the Sun are considered ejected by Jupiter. For simplicity, the same condition is applied when the comet is close to the interloping star, assuming it is orbited by a similar giant planet.

Numerical integrations are performed with the 15th-order RADAU integrator of \citet{everhart1985}. With the limited time span of our integrations compared to the time step used, this integrator is perfectly suitable. It is not energy preserving (symplectic), but for the problem at hand, small departures from energy conservation are not an issue, and the integrator has been found to perform very well in closure tests \citep{carusi1985}.

\section{Results} \label{sec:results}

We perform a series of 1\,000 Monte Carlo simulations for each cluster, differing from each other in terms of the initial solar orbit and cometary orbits. The total number of comets treated is thus $3 \times 10^6$, for each cluster. 
With such statistical sampling, we shall present not only the typical behaviours, but also the tail of the probability distribution, \ie, the rare outcomes. Obviously, in the latter cases, our results are not statistically accurate but rather indicative.

\subsection{Comet survival probability}

\begin{figure*}
\centerline{
	\includegraphics[height=7.5cm]{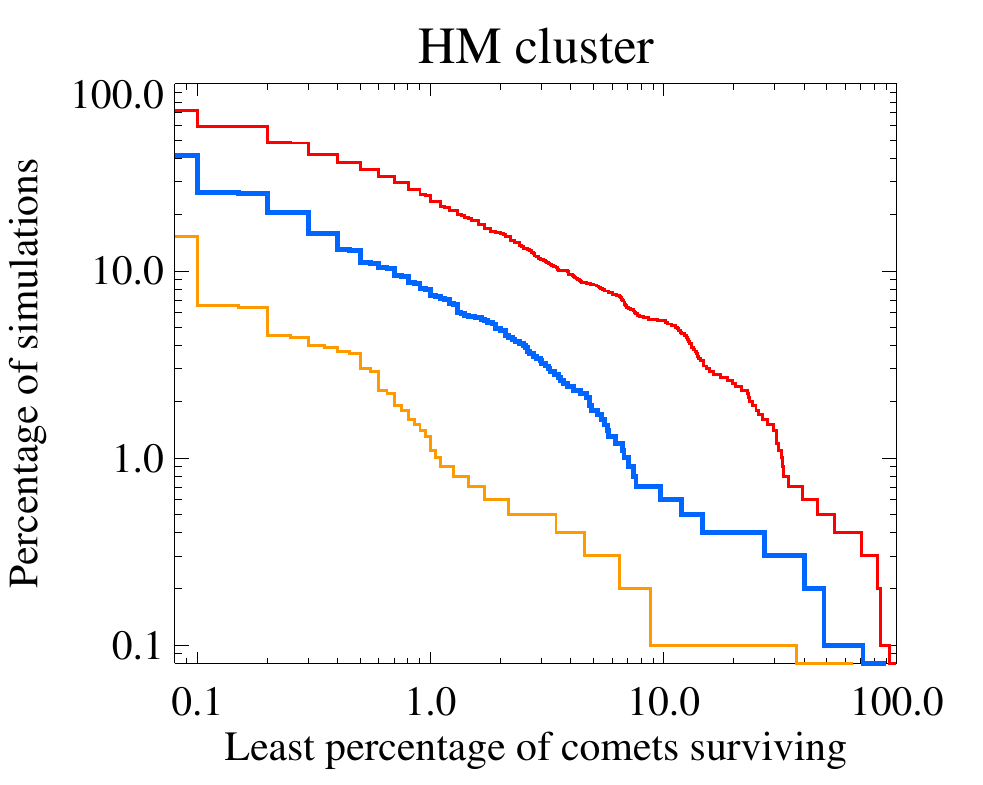}
	\includegraphics[height=7.5cm,clip,trim=2em 0 0 0]{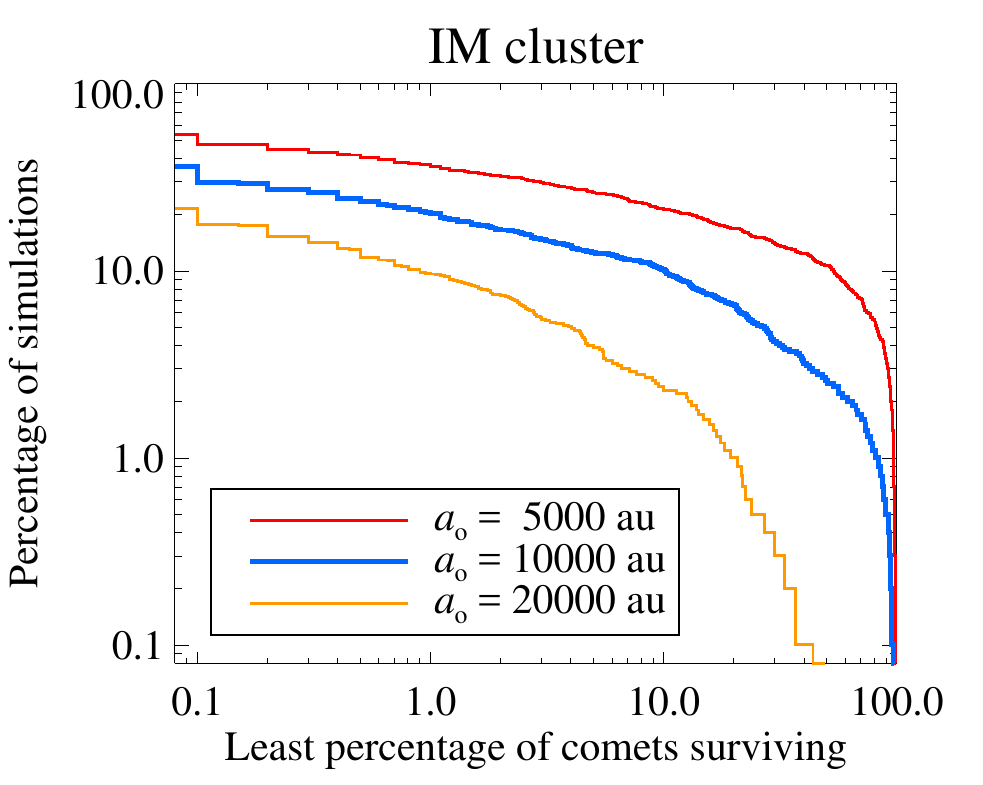}
}
\caption{Distribution of comet survival probability for the three different bins of initial semimajor axis, $a_{\rm o} = 5000$\unit{au} (red), $10\,000\unit{au}$ (blue), and $20\,000\unit{au}$ (orange). 
\textit{Left panel:} The high-mass cluster. 
\textit{Right panel:} The intermediate-mass cluster. 
}
\label{fig:survival}
\end{figure*}

We illustrate in Fig.~\ref{fig:survival} the statistical distribution of comet survival probability in these simulations. The term ``survival'' refers to those comets that do not experience ejection either by moving to distances exceeding 1\unit{pc} or by intruding into the loss cone -- hence they stay in the Oort Cloud until the end of the integration. Since it is natural to imagine that stellar encounters cause comets to leave the cloud directly, while the cluster tide may perturb their orbits into the loss cone, we have also performed two extra sets of 100 simulations each, where we artificially turned off the cluster tide and the stellar encounters, respectively. By comparing these results with those of the full model, we hope to learn which is the dominant mechanism causing the loss of comets from the Oort Cloud.

Results for the HM cluster are presented in the left panel of Fig.~\ref{fig:survival}. In this case, the median survival probability for comets in the initial sub-population representing the primordial \inner\ comet cloud, $a_{\rm o} = 5\,000\unit{au}$, is just 0.2\unit{\%}, representing a few comets per simulation. 
For comets representing the primordial \intermediate\ and \outer\ comet clouds, $a_{\rm o} = 10\,000\unit{au}$ and $20\,000\unit{au}$, the median survival probability is less than 0.1\unit{\%}, \ie, fewer than one comet per simulation.
Hence, typically the entire primordial outer comet cloud is lost. In only 1.3\unit{\%} of simulations do more than 1\unit{\%} of the primordial outer cloud members survive.
For the primordial inner and intermediate comet cloud, more than 5\unit{\%} of comets survive in 9\unit{\%} and 2\unit{\%} of the simulations, respectively.

Comparing now with the results for the IM cluster in the right panel of Fig.~\ref{fig:survival}, we see an important difference. The curves representing the fall-off of the percentage of simulations with increasing percentage of survivors are much flatter for the IM cluster over almost the whole range. Consequently, the median survival probability for the primordial \inner\ comet cloud is only 0.1\unit{\%}, \ie, even lower than for the HM cluster, while the percentage of simulations with a much larger number of survivors is considerably higher for the IM cluster. For instance, the fraction of simulations with more than 1\unit{\%} survivors in the primordial outer cloud is 9.8\unit{\%} for the IM cluster, compared to 1.3\unit{\%} for the HM cluster. On the other hand, the escape times of comets from the Sun in the two clusters are much shorter in the IM than in the HM case. For the inner, intermediate and outer clouds, the median escape times are 12, 5 and 4\unit{Myr}, respectively, in the IM cluster while in the HM cluster these are 51, 47 and 20\unit{Myr}.

The reason for these differences has to do with the typical fate of the Sun in the two clusters. As we shall see in Sect.~\ref{sec:solar_orbit}, 
after 400\unit{Myr} the Sun typically remains in the HM cluster and escapes from the IM cluster. 
Specifically, the \textit{remaining} percentage is 94\unit{\%} in the HM case and the \textit{escaping} percentage is 95\unit{\%} in the IM case. This means that the behaviours exhibited in the two panels of Fig.~\ref{fig:survival} may actually carry as much information about whether the Sun remains in the cluster, as the difference between HM and IM clusters.

Although the statistics is rather poor for the less common situations, we can still compare the fate of Oort Cloud comets in all four cases, \ie, HM vs IM clusters and remaining vs escaping Sun. We have thus found that for the IM cluster there is not much difference of comet survival statistics, whether the Sun stays in the cluster or it escapes. For the HM cluster the fate of the comets is more sensitive to the fate of the Sun. When the Sun escapes from this cluster, the comet survival statistics is intermediate between the 
left and right panels of Fig.~\ref{fig:survival}.

Our results are consistent with the following picture. As seen in Fig.~\ref{fig:encounter_frequency}, the central part of the IM cluster starts out with a very high encounter frequency, but this falls off rapidly with time due to cluster evaporation. The trend is similar for the strength of the cluster tide. Thus, during the early phases the solar system runs a high risk of being stripped of its entire Oort Cloud due to very close stellar encounters, if the solar orbit penetrates close to the centre, but our simulation only covers a few of the strongest encounters expected. As we shall see below, the cluster tide also plays a role in this context. We may therefore expect to see a majority of disastrous cases with no or very few comets surviving and at the same time another category of cases, where the most perilous encounter was weaker and left a significant part of the Oort Cloud bound to the Sun. This situation quickly became fossilized, as the cluster started to dissolve and the Sun migrated outward before finally escaping.

Consequently, the time of escape might not matter very much for the survival statistics of the Oort Cloud, and even in case the Sun remains in the IM cluster for the whole interval considered, relatively little further damage to the cloud may be the rule. The case of the HM cluster is different, because its central region remains perilous for the full length of the integrations. Therefore, in case no disastrous encounter occurs during the early stage, the remaining part of the Oort Cloud will in general be subject to further damage due to close encounters. The chance for a significant fraction of surviving comets is relatively small. However, there are of course situations, where the Sun undergoes efficient outward migration, and the survival rate is higher. This will be the case in particular for simulations where the Sun escapes from the cluster.

As shown in Table~\ref{tbl:meddist1}, the comparison simulations where we artificially turned off either the cluster tide or the stellar encounters, indicate 
that the survival probability for the HM cluster is hardly affected at all by the presence or absence of the cluster tide in the dynamical model, while the removal of the stellar encounters drastically increases the chances of survival for all the primordial cloud populations. It is thus clear that the losses of comets are mainly due to the stellar encounters. However, for the outer cloud we also note that the median survival probability is very low even without stellar encounters. Hence, in this part of the cloud -- and only in this part -- the comets are destabilized also by the cluster tide.

We also see an indication in Table~\ref{tbl:meddist2}  that, for the IM cluster, both the cluster tides and the stellar encounters matter for the loss of comets from the Oort Cloud -- the tides even more than the encounters. Here we have to take note of the fact that the scenario in the model without stellar encounters is very different from that of the other two models. In the full dynamical model the Sun escapes from the cluster in a large majority of cases, as seen above, and the same is true for the model without the cluster tide. However, when there are no stellar encounters, the solar orbit remains practically unchanged -- hence, the Sun never escapes but stays close to its initial orbit during the whole integration. The initial orbits penetrate close to the cluster centre and the comets are hence sensitive to the radial tide. This explains the extensive losses of comets from all parts of the Oort Cloud.

\subsection{Comet end states}

In Tables~\ref{tbl:meddist1} and \ref{tbl:meddist2} we present statistics regarding the three end states, for both clusters: the direct departures leading to unlinking of the comets, the entries into the loss cone leading eventually to hyperbolic ejection by Jupiter, and the survivals until the end of the integration. These are shown for the full dynamical model as well as the two comparison models, and for each set of primordial comet orbits. In each case, the listed percentages refer to the median of the simulations.

The most striking feature concerning the HM cluster (Table~\ref{tbl:meddist1}) is the predominance of direct departures in the full model as well as the model without tides. The model without encounters is different as regards the inner and intermediate comet populations, where instead of the predominance of departures we find large fractions of survival or loss cone intrusion. Once more we see that the outer population is very vulnerable to direct departures even without encounters. Apparently, the cluster tide causes a strong instability of the outer cloud orbits but much less so for the orbits of the other parts.

The loss cone entries practically only appear in the presence of the cluster tide, showing that stellar encounters very rarely cause such an evolution. On the contrary, stellar encounters are seen to interfere with the tidal evolution of the perihelion distance, preventing the loss cone entries from the inner and intermediate cloud that would otherwise occur. By plotting the perihelion distance vs time in the tide-only model, we have verified that the loss cone entries are caused by a secular oscillation of orbital angular momentum driven by the cluster tide. Even though such a pulsation is also a feature of the Galactic disk tide currently experienced by Oort Cloud comets \citep{heisler1986}, the dynamics inside the cluster is basically different. The cluster tide is radial and non-conservative. The amount of the energy exchange is shown by the preference for tidally caused departures of Oort Cloud comets belonging to the outer population.

Comparing with the results for the IM cluster (Table~\ref{tbl:meddist2}), the main difference appears for the model without stellar encounters. As noted above, the IM model is special in that the Sun remains more or less locked to its initial orbit for the full length of the integration. 
As discussed in Sect.~\ref{sec:solar_orbit}, 
these orbits tend to have pericentre distances less than 0.5\unit{pc}. That this exposes the Oort Cloud comets to a very strong cluster tide can be realized from Fig.~\ref{fig:cluster_structure}, because the strength of the cluster tide is proportional to the mass density in the homogeneous, central part. This density is seen to be very high at all times in the IM cluster, and the region of homogeneity extends to $r\simeq 0.1$\unit{pc}.

For this reason we see a very high fraction of unlinked comets in all parts of the Oort Cloud. Of course, the model without stellar encounters is not meant to be realistic. In reality the stellar encounters would rapidly change the solar orbit, thereby in general decreasing the influence of the cluster tide. What this model shows is that in an IM cluster the central region is very dangerous for Oort Cloud comets due to the cluster tide, and in case the solar orbit remains with a small pericentre distance for too long, the cloud is likely to be stripped away.

\begin{table} \centering
\caption{Median probabilities (\%) of comet end states in models of the high-mass cluster with/without cluster tide and stellar encounters.} 
\label{tbl:meddist1}
\begin{tabular}{r rrr} \hline\hline \noalign{\smallskip} 
\multicolumn 1 c {$a_0$} & \multicolumn 1 c {Unlinked} & \multicolumn 1 c {Loss cone} & \multicolumn 1 c {Survived}\\
\multicolumn 1 c {(au)} \\
\hline \noalign{\smallskip} \multicolumn 4 c {Full dynamical model\tablefootmark a} \\ \hline \noalign{\smallskip} 
   5000 &   95.3 &    4.0 &    0.2 \\
   10\,000 &   97.2 &    2.7 &    0.0 \\
   20\,000 &   99.6 &    0.4 &    0.0 \\
   All &   97.7 &    2.1 &    0.0 \\
\hline \noalign{\smallskip} \multicolumn 4 c {No cluster tide\tablefootmark b} \\ \hline \noalign{\smallskip} 
   5000 &   98.7 &    0.9 &    0.3 \\
   10\,000 &   99.6 &    0.3 &    0.0 \\
   20\,000 &   99.9 &    0.1 &    0.0 \\
   All &   99.6 &    0.3 &    0.0 \\
\hline \noalign{\smallskip} \multicolumn 4 c {No stellar encounters\tablefootmark b} \\ \hline \noalign{\smallskip} 
   5000 &    0.0 &   16.6 &   83.4 \\
   10\,000 &   13.4 &   18.5 &   65.2 \\
   20\,000 &   98.2 &    0.6 &    1.1 \\
   All &   13.4 &   13.6 &   65.2 \\
\hline \noalign{\smallskip}
\end{tabular}
\tablefoot{
\tablefoottext a {Based on 1000 simulations.}
\tablefoottext b {Based on 100 simulations.}
}
\end{table}

\begin{table} \centering
\caption{Same as Table~\ref{tbl:meddist1}, but for the intermediate-mass cluster.} \label{tbl:meddist2}
\begin{tabular}{r rrr} \hline\hline \noalign{\smallskip} 
\multicolumn 1 c {$a_0$} & \multicolumn 1 c {Unlinked} & \multicolumn 1 c {Loss cone} & \multicolumn 1 c {Survived}\\
\multicolumn 1 c {(au)} \\
\hline \noalign{\smallskip} \multicolumn 4 c {Full dynamical model\tablefootmark a} \\ \hline \noalign{\smallskip} 
   5000 &   97.5 &    1.9 &    0.1 \\
   10\,000 &   99.6 &    0.3 &    0.0 \\
   20\,000 &   99.9 &    0.1 &    0.0 \\
   All &   99.5 &    0.4 &    0.0 \\
\hline \noalign{\smallskip} \multicolumn 4 c {No cluster tide\tablefootmark b} \\ \hline \noalign{\smallskip} 
   5000 &   98.5 &    0.3 &    1.1 \\
   10\,000 &   99.6 &    0.1 &    0.2 \\
   20\,000 &   99.8 &    0.1 &    0.0 \\
   All &   99.5 &    0.2 &    0.2 \\
\hline \noalign{\smallskip} \multicolumn 4 c {No stellar encounters\tablefootmark b} \\ \hline \noalign{\smallskip} 
   5000 &   96.5 &    2.9 &    0.5 \\
   10\,000 &   99.4 &    0.5 &    0.0 \\
   20\,000 &   99.9 &    0.0 &    0.0 \\
   All &   99.3 &    0.5 &    0.0 \\
\hline \noalign{\smallskip}
\end{tabular}
\tablefoot{
\tablefoottext a {Based on 1000 simulations.}
\tablefoottext b {Based on 100 simulations.}
}
\end{table}

\begin{figure*}
\centerline{
	\includegraphics[height=7.5cm]{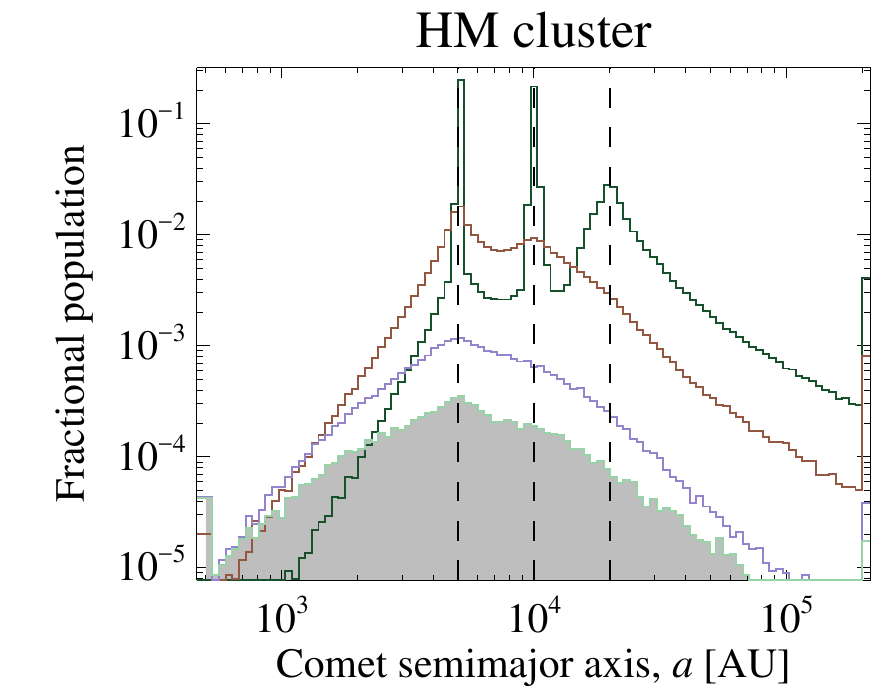}
	\includegraphics[height=7.5cm,clip,trim=3em 0 0 0]{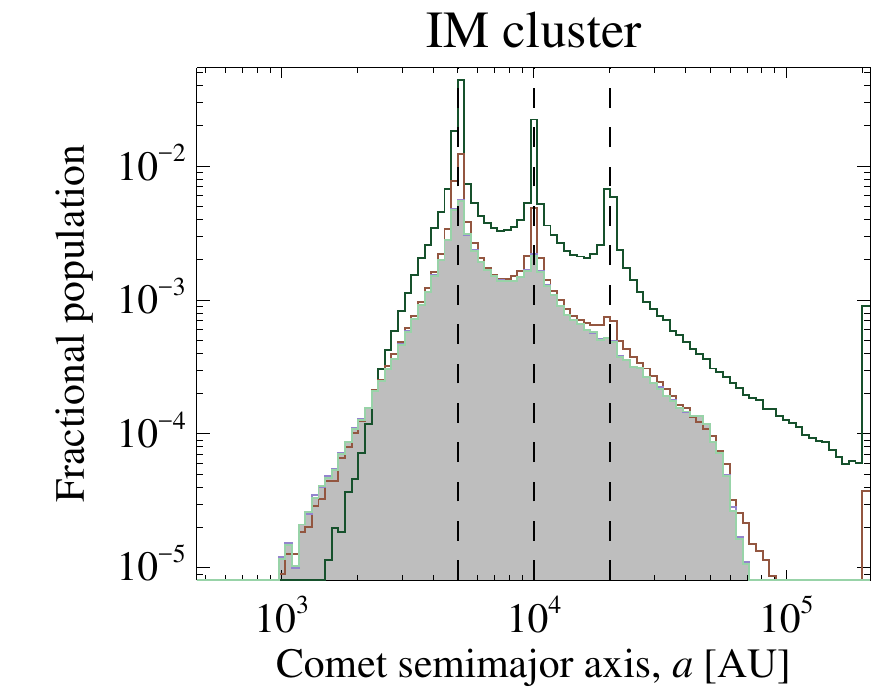}
}
\caption{Time evolution of the combined comet population of all simulations with the full dynamical model. The fractional populations are counted with respect to the total initial population. Results were sampled at 10\unit{Myr} timesteps, using 100 uniform logarithmic bins for $a \in [500,200\,000]\unit{au}$. The locations of the three initial sub-populations are indicated by vertical dashed lines. 
Here, snapshots are shown at times $T = 10$, 50, 200, and 400\unit{Myr}, from top to bottom, using black, red, blue and green colours, respectively, where the final time step is also shaded in grey. Comets beyond the plot limits have been summed at the edges. 
\textit{Left panel:} The high-mass cluster. 
\textit{Right panel:} The intermediate-mass cluster. 
The three remaining peaks in the final distribution reflect the fact that the vast majority of the stars in this cluster model have by then left the cluster with fossilized structures of their cometary clouds.
}
\label{fig:evolution}
\end{figure*}

\subsection{Oort Cloud evolution}

The time evolution of the comet cloud is illustrated in Fig.~\ref{fig:evolution}. We will first discuss the left panel, showing the case of the HM cluster. 
The outermost population is typically dispersed by the very first close stellar encounter in each simulation, giving rise to a continuous distribution of semi-major axes reaching beyond 1\unit{pc}. Such wide orbits are  unstable in the current Galactic environment and even more so in a dense cluster. The comets are rapidly lost by the two energy-perturbing agents identified above, \ie, the stellar encounters and the cluster tide.

After some 200\unit{Myr}, the primordial populations have dispersed sufficiently that the comet orbits appear rather smoothly distributed, albeit retaining a broad central peak covering $a \sim 2\,000$--$15\,000\unit{au}$. The subsequent time evolution sees the outer population, $a > 10\,000\unit{au}$, diminish at a rate similar to the central population.
This means that a steady state is reached, whereby this outer population remains as a transit stage of comets migrating from the inner parts and eventually departing from the solar system due to energy perturbations. The semi-major axis distribution for $a > 20\,000\unit{au}$ is seen to evolve toward a power-law slope of $-2$ corresponding to a flat energy distribution, which is typical of a diffusion process with an absorbing wall near $1/a = 0$ caused by our definition of departures.

The innermost part of the comet cloud, $a < 5\,000\unit{au}$, is rapidly populated during the first 50\unit{Myr}. The core of this population, $a \lesssim 3\,000\unit{au}$, then remains essentially inert, being depopulated only in those individual simulations where very strong and close encounters occur. 
This core appears to form as a rule rather than an exception, comprising roughly 20\unit{\%} of the surviving comets from the primordial \inner\ cloud, while the primordial \intermediate\ and \outer\ clouds each contribute one and two orders of magnitudes fewer comets (see below). The core is thus more than twice as populated as the outward migrators beyond 20\,000\unit{au} at the time ($T = 400\unit{Myr}$), when we stop the integrations. 

From the plotted results, we also have an indication about what would happen, if we had continued the integration further in time. The peak would remain close to $5\,000\unit{au}$, and the curve would flatten out at smaller semi-major axes, while it would continue to be shifted downward at larger semi-major axes. Thus the predominance of the quasi-inert core would become further accentuated, as the total population of the cloud continues to decrease.

The evolution of the cloud in the IM cluster case -- shown in the right panel -- is basically similar, but some differences are easily seen. After 50\unit{Myr} the structure undergoes very small changes. As a rule, the Sun has then left the cluster or migrated out of the central region for most of the time. We noted above that this leads to a fossilized structure of the cloud, which we here see represented by the histograms in red, blue and green. The inner core is less pronounced than in the HM case. The outermost part of the cloud is cut at $a\sim 50\,000$\unit{au}, since there are no more perturbations large enough to replenish these orbits from the inside.

\begin{figure*}
\centerline{
	\includegraphics[height=8cm]{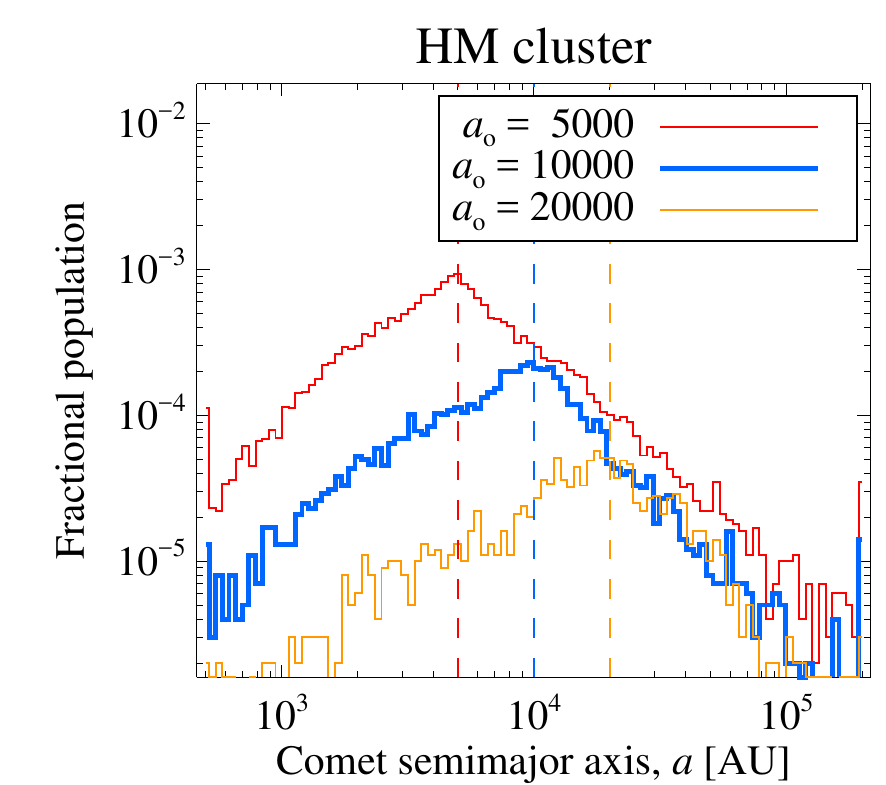} 
	\includegraphics[height=8cm, clip,trim=3em 0 0 0]{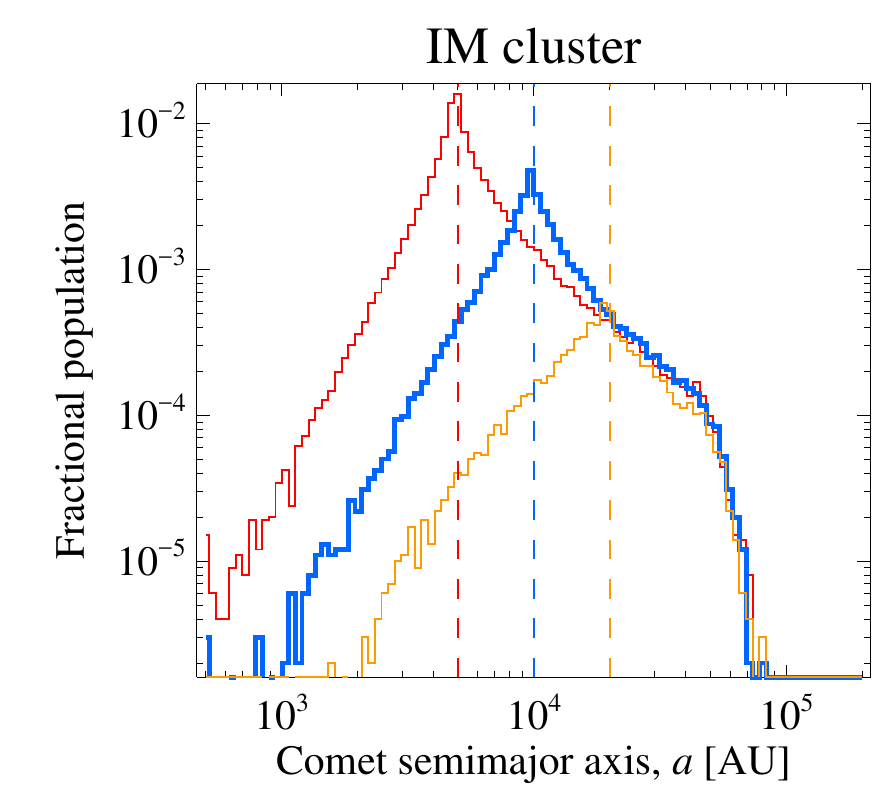}
}
\caption{Mean semi-major axis distributions of the surviving comets at the end of the integrations, combined from all simulations, for the three primordial populations as in Fig.~\ref{fig:survival}, using the same colours and styles of the curves. Vertical dashed lines represent the initial values. Differential distributions are shown by histograms, using a common normalization to the size of the initial populations. The curves are shaped by energy perturbations caused by the external agents. Statistical noise is seen mainly for the inner and intermediate primordial populations.
\textit{Left panels:} The high-mass cluster. 
\textit{Right panel:} The intermediate-mass cluster. 
}
\label{fig:dist}
\end{figure*}

We illustrate in Fig.~\ref{fig:dist} the separate mean semi-major axis distributions for the surviving comets of the three primordial populations. The HM cluster is shown to the left and the IM cluster to the right. 
The median values of the survivors in the HM cluster are less than the initial values -- 4\,510, 8\,000 and $15\,700\unit{au}$ for the inner, intermediate and outer populations, respectively. These shifts are due to the preferential loss of comets reaching large semi-major axes and the relative safety of comets diffusing toward smaller values. The shifts are smaller in the IM cluster, especially for the inner and central populations. This is likely due to the absence, in most cases, of a long-term energy diffusion.

Thanks to the common normalization used, the fact that the outer population has the smallest number of survivors in both clusters is clearly displayed in Fig.~\ref{fig:dist}. The mean survival probabilities in the HM cluster are 2.1, 0.6, and 0.1\unit{\%} for the three respective populations. In the IM cluster these values are 11.7, 4.1, and 0.8\unit{\%}. 
All these values are significantly larger than the corresponding medians (Tables~\ref{tbl:meddist1} and \ref{tbl:meddist2}), because there is significant spread between the results of different simulations, and the survivors are concentrated to the minority that had the smallest external effects.

The sums of the three differential distributions yield the distributions shown at $T = 400\unit{Myr}$ in Fig.~\ref{fig:evolution}. Each of these has a roughly triangular shape in the log-log diagrams used with a maximum at the initial value of the semi-major axis. For the HM cluster we see a steeper slope for larger than for smaller values. As noted above, the steeper slope is close to $-2$, and the core population created by inward migration has contributions differing by roughly one order of magnitude between the inner, intermediate and outer primordial populations. For the IM cluster the distributions are more symmetric around the maximum until the cut at large semi-major axes is reached. The slopes are higher on both sides of the maximum than in the HM cluster case.

\begin{figure*}
\centerline{
\includegraphics[height=7.5cm]{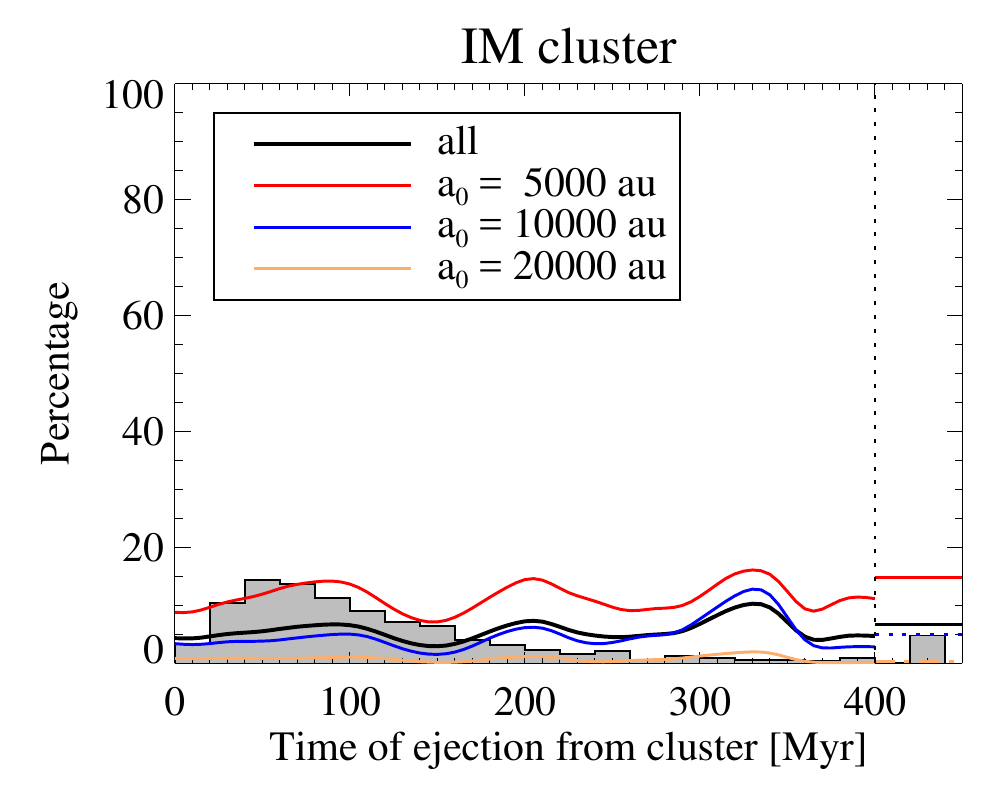}
\includegraphics[height=7.5cm,clip,trim=3em 0 0 0]{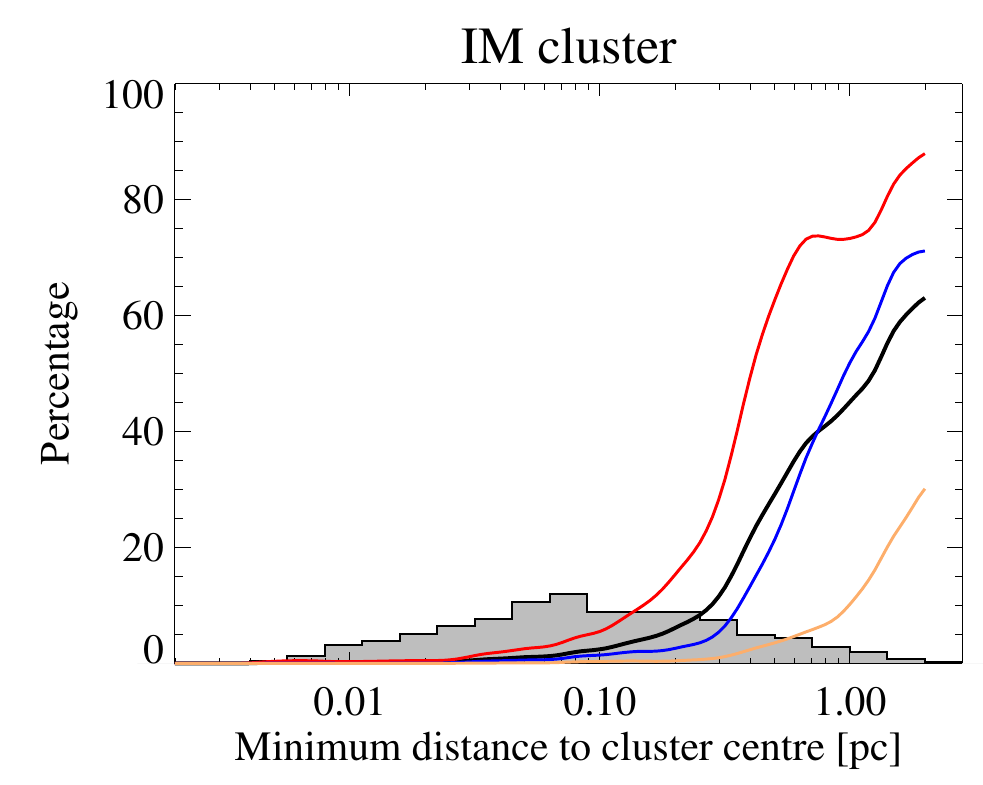}
}
\caption{
Statistics for the IM cluster over the distribution of simulations (grey-shaded histogram) and the survival percentages of comets.
The comet populations are shown both combined (thick black curve) and for each population (coloured curves). The survival statistics are computed as the arithmetic mean using a boxcar, and then smoothed by a Gaussian kernel.
\textit{Left panel:} Statistics as a function of the time when the Sun is ejected from the cluster, using a boxcar width of 50\,Myr and Gaussian kernel $\sigma = 10$\,Myr. Cases where the Sun is not ejected from the cluster are shown to the right of the vertical dashed line.
\textit{Right panel:} Statistics as a function of the Sun's nearest distance to the cluster centre, computed in logarithmic bins using a boxcar width of 0.3\,dex and Gaussian kernel $\sigma = 0.05$\,dex.
}
\label{fig:ejtime_mindis}
\end{figure*}

We have already noted a few special features of the evolution experienced within the IM cluster. One is that the Sun tends to leave the cluster during our simulations, whereby the influence of the cluster on the Oort Cloud is terminated. The other is that the IM cluster is equipped with a high-density central core, which acts as a very efficient pitfall to Oort Cloud comets, in case the solar orbit enters into its vicinity. To explore the influence of these features on the survival of the Oort Cloud, in Fig.~\ref{fig:ejtime_mindis} we illustrate the relevant statistical properties: histogram distributions of the time when the Sun is ejected from the cluster and the minimum periapsis distance of the solar orbit, and together with these the variations of the Oort Cloud average survival probability with the parameters in question.

The most striking fact revealed by the two panels of Fig.~\ref{fig:ejtime_mindis} is that the minimum periapsis distance effectively governs the Oort Cloud survival probability, while the time of ejection of the Sun does not exhibit any similar influence. It is clear that any approach of the Sun to less than $0.2-0.3$\unit{pc} from the cluster centre leads to the loss of almost the entire Oort Cloud including the comets we refer to as the inner cloud. On the other hand, for periapsis distances approaching 1\unit{pc} most of the inner and intermediate parts will survive. We interpret this to reflect the tidal action of the cluster core, whose radius is approximately $0.1-0.2$\unit{pc} (see Fig.~\ref{fig:cluster_structure}). Since the median of the minimum periapsis distance is seen to be close to 0.1\unit{pc}, such tidal losses can fully explain the generally low survival probability of Oort Cloud comets in the IM cluster.

Of course, when the Sun penetrates into or close to the cluster core, it can also experience close stellar encounters that strip comets away from the Oort Cloud. We showed in Table~\ref{tbl:meddist2} that the median survival probability is very low, both if we turn off the cluster tide, and if we turn off the stellar encounters. This demonstrates that stellar encounters do have an influence. However, the tidal effect is probably the dominant effect. In our model, only one star will be closely encountered during each periapsis passage of the Sun, and it seems unlikely that the orbit with the smallest distance from the cluster centre will invariably involve an encounter that is efficient enough to strip away almost the whole cometary cloud. Thus, some scatter would be expected, making the survival curves less smooth and monotonic if stellar encounters were dominant.

The wavy pattern exhibited by the survival probability with respect to time of ejection reflects the limited statistics. Although the curves are smoothed, their maxima are strongly influenced by the occasional simulations, where the minimum periapsis distances are large and many comets survive. Ejection times close to that of one of these simulations will generate a relatively large survival probability in the calculated distribution. The waves for different initial cloud populations are in phase, because these groups are simulated with identical solar orbits and encounters. The absence of any systematic decrease of the survivability with time of ejection merely shows that the Sun may spend a long time in the cluster with its Oort Cloud intact, provided that it does not penetrate into the cluster core.

\subsection{The solar orbit}
\label{sec:solar_orbit}

\begin{figure*}
\centerline{
	\includegraphics[height=15.8cm]{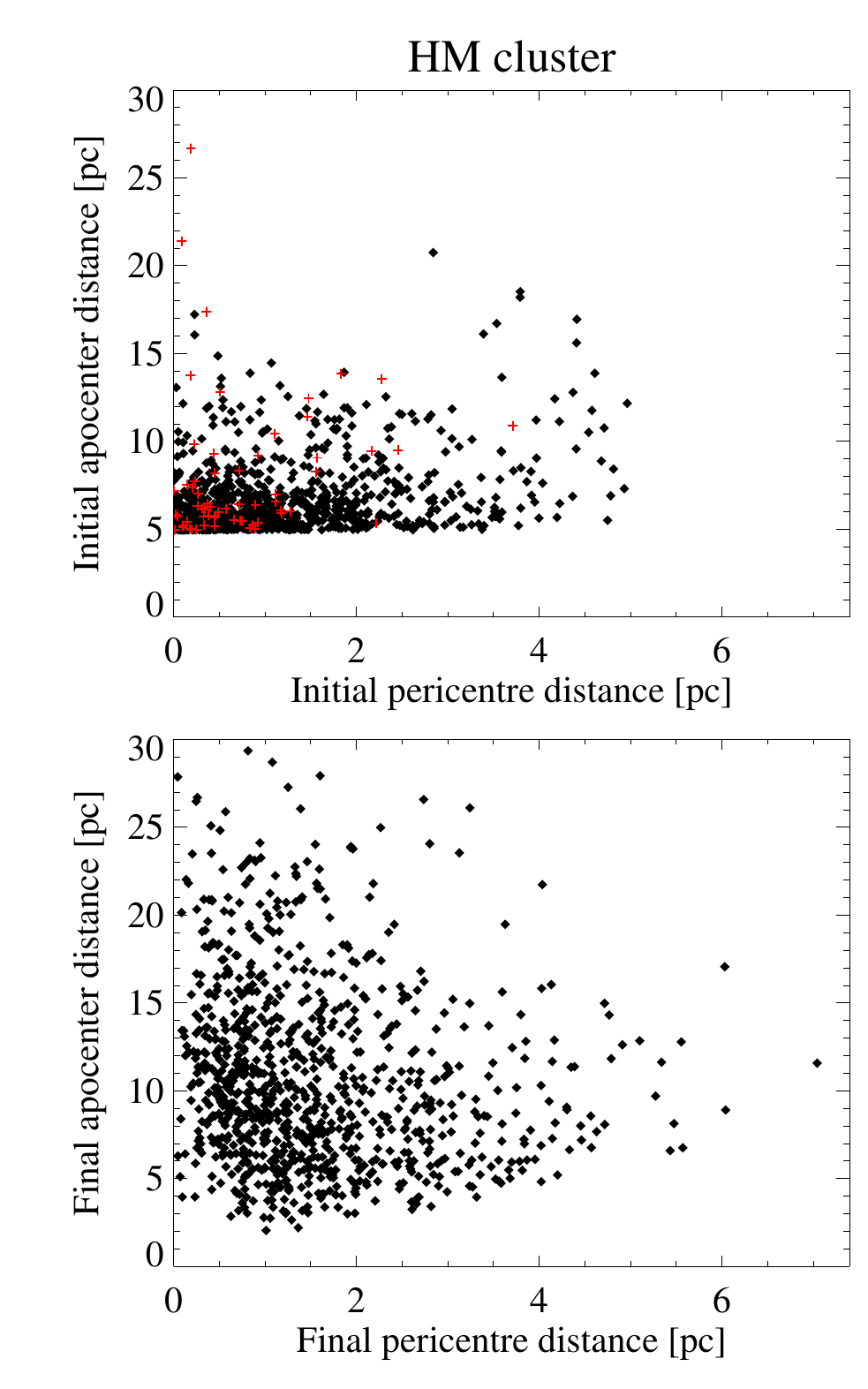}
	\includegraphics[height=15.8cm,clip,trim=4em 0 0 0]{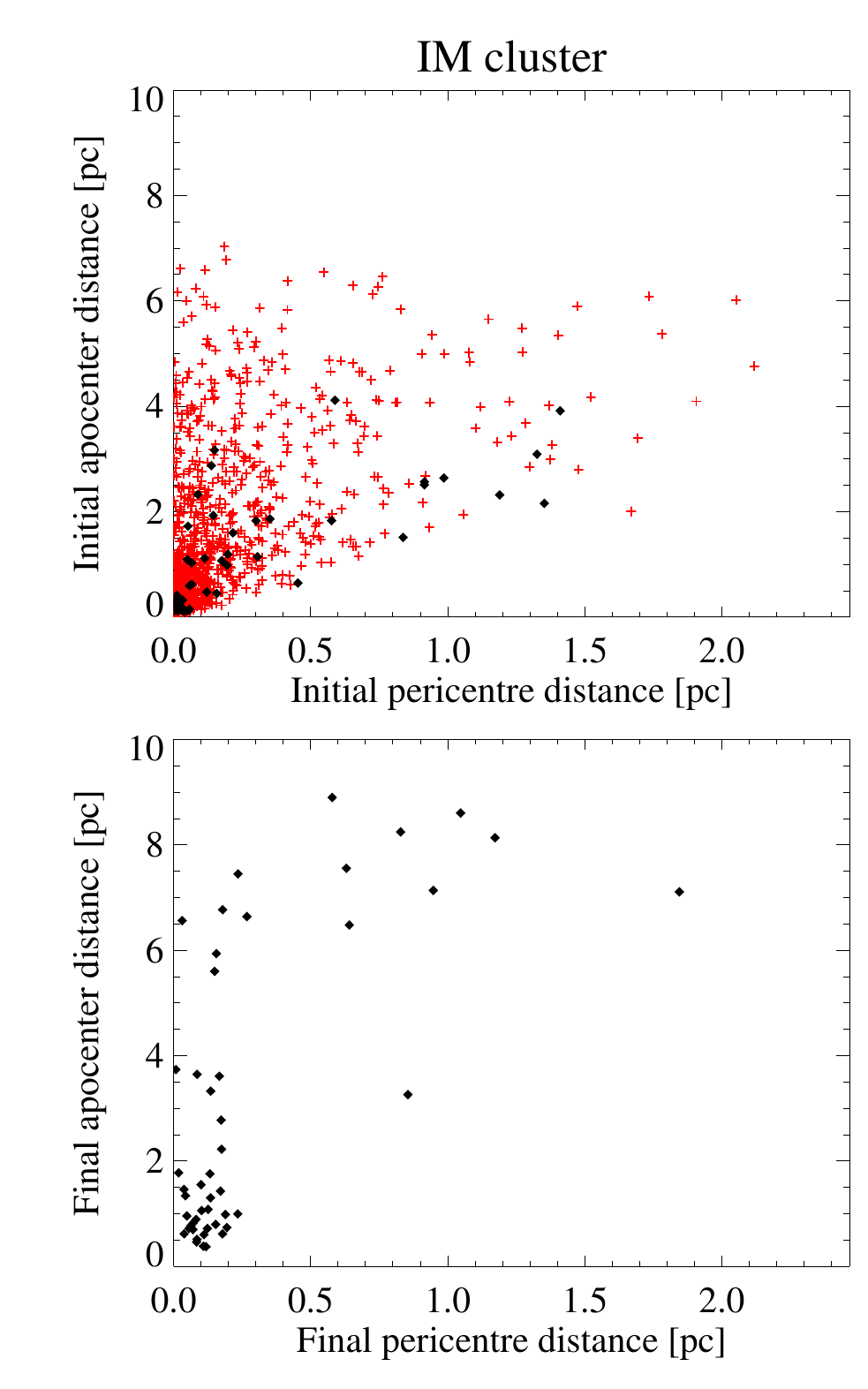}
}
\caption{Distributions of initial and final solar orbits for all our simulations using the full dynamical model. Pericentre ($q$) and apocentre ($Q$) distances of the rosette orbits are plotted on the axes of each diagram.
\textit{Left panels:} The high-mass cluster. 
\textit{Right panels:} The intermediate-mass cluster. 
\textit{Top panels:} Initial orbits. For the HM cluster, these orbits start from the half-mass radius ($r=4.87\unit{pc}$) with some random radial velocity, so we always have $q < 4.87\unit{pc} < Q$. The red crosses denote orbits, where the Sun was ejected from the cluster during the simulation, while the rest are denoted by black diamonds.
\textit{Bottom panels:} Final orbits. Only orbits where the Sun survived as a cluster member are shown.
}
\label{fig:sorb}
\end{figure*}

Let us finally present some results on the evolution of the solar orbit in the cluster, caused by the impulses received from the encountering stars. Figure~\ref{fig:sorb} shows two scatter diagrams of apocentre vs pericentre distances at the beginning of our integrations (upper panels) and at the end (lower panels). The left pair of panels refers to the HM cluster and the right pair illustrates the IM cluster. Each symbol represents one of the 1\,000 simulations with the full dynamical model. The red symbols in the initial distributions mark those solar orbits which were not stable, so that the Sun was ejected from the cluster before the end of the integrations (this fate was registered, if the Sun moved beyond the tidal radius, given in Table~\ref{tbl:modelparams}, in each cluster).

The cases of ejection amount to 5.8\% of the simulations for the long-lived, HM cluster. This verifies the expectation for the Sun as a relatively massive cluster star that it suffers only a small risk of ejection within 400\unit{Myr}. For the IM cluster the situation is the opposite. The cluster as a whole dissolves on a much shorter time scale and the Sun is no exceptional star. We find that the median survival time of the Sun as a cluster member is about 100\unit{Myr}, while in 80\% of the simulations the Sun is ejected within 200\unit{Myr}. After 400\unit{Myr}, the Sun remains in only 4.8\% of the simulations.

For both cluster models, the initial and final distributions of solar orbits differ markedly from each other. Most of the time the Sun is pushed outwards in the cluster, since both apocentre and pericentre distances show  increasing trends -- more pronounced for the apocentre distance. For a minority of cases in the HM cluster, the solar orbit evolves into a smaller apocentre distance or a lower effective eccentricity. In the IM cluster, most of the remaining solar orbits stay close to the cluster centre.

We caution that the trend toward the outskirts of the HM cluster may be affected by our manner of selecting the stellar encounters. Choosing each time the encounter with the largest strength parameter (Sect.~\ref{sec:encounterselect}) means that the massive stars are favoured as encounter partners. Statistically, during binary encounters, energy per unit mass flows from the more massive to the less massive partner -- a basic reason for mass segregation causing the concentration of high-mass stars to the cluster core and the preferential escape of low-mass stars \citep{Spitzer1969}. Had we included all the encounters, in which case the simulations had been more realistic, the outward migrating trend for the Sun would likely have been balanced by many interactions with low-mass stars and thus reduced.

Another word of caution is justified concerning the IM cluster. Here, again, our result may be biased by our modelling of the cluster. In reality, the central region of the cluster becomes enriched in massive stars, but we neglected this mass segregation. Thus, our encounter selection -- while preferring massive partners -- did not favour these as much as it should have done in a realistic modelling. This may have made it easier for the Sun to stay in the central region rather than be expelled from it.



\section{Discussion} \label{sec:discus}
\subsection{Modelling issues} \label{sec:modiss}

We have chosen to base one of our cluster models on M67, which according to \citet{hurley2005} started out with more than 20\,000 stars. This was done in spite of the conclusion by \citet{adams2010} that a birth cluster with more than about 10\,000 stars would threaten the stability of the planetary orbits. One reason not to worry is that Adams considered giant planets on their current orbits, while the Nice Model holds that the orbits of the giant planets had much smaller semi-major axes during the early epochs of solar system history -- both in its original form \citep{tsiganis2005} and in later versions. This would obviously reduce their vulnerability to external perturbations. However, the intricate resonant clockwork of the Nice Model \citep{Levison2011} might be upset by stellar encounters far more distant than those previously considered as disastrous. No analysis of this problem has yet been made to our knowledge.

We did not aim to survey the full spectrum of cluster sizes but just took two examples of relatively rich systems that are statistically likely birthplaces of solar type stars \citep{Lada2003}. We found that the intermediate-mass cluster is in a certain sense as hostile to the survival of a primordial Oort Cloud as the high-mass one. However, it is dangerous to extrapolate this to even smaller birth clusters. At some point, the cluster dissolves so quickly that the destructive influence on the Oort Cloud disappears. It remains to find out at which initial cluster mass this transition occurs.

Meanwhile, it is worth noting that a surviving Oort Cloud formed in a very dense cluster environment should be be very tight including semi-major axes far smaller than those that we have investigated \citep{FernBrun2000,brasser2006}. We have seen that such an inner core does survive in the clusters we modelled. An important issue is then how to activate this core and repopulate the outer halo after the star leaves the cluster. In a sense, a fossilized inner core would be ineffective: it would not provide any observable comets. We shall return to this discussion in Sect.~\ref{sec:conclu}.

As to the Sun, an active Oort Cloud was formed in the simulations by \citet{levison2010}, who considered a very small and short-lived birth cluster. In fact, judging from \citet{adams2010}, many of the relevant cluster sizes would not be in absolute conflict with the nucleosynthesis requirement, since they would yield a random probability of $\sim 10$\% for a relevant supernova explosion.

Another issue concerns the realism of our dynamical model for the external perturbations suffered by the Oort Cloud comets. Of course, only a full $N$-body simulation of the cluster may be considered fully realistic, but this is beyond the scope of our preliminary study. The synthetic model that we developed represents the cluster by its smooth potential field plus a sequence of two-body encounters within this field, which the Sun experiences with individual cluster stars. This is quick and efficient, but it certainly departs from reality. We have assumed the Sun and the comets to experience, in addition to the smooth cluster potential, the fluctuating component caused by just one passing star at a time. In reality there are typically several nearby stars contributing to this fluctuating field at the same time. Hence, concerning the loss of comets, in many cases the fate of a comet may critically depend on the details of the tidal field, so that our approximation may either save comets from leaving the Sun or stimulate their escape, depending on the circumstances.

Although it does not seem likely that any serious systematic errors occur as a result of our simplifications, a full answer cannot be found except through time-consuming N-body simulations. One technique would be to relax the blocking of overlapping encounters to see, if by allowing up to 5--10 simultaneous encounters and treating these $N$-body systems accurately, we would get statistically different results on the loss of comets. However, this would mean an additional major effort, which still would not solve all problems.

\subsection{Simplifying assumptions} \label{sec:simpli}

The most important simplification that we used may be the assumption of a step function for the initial population of the Oort Cloud. In fact,  the Oort Cloud was not built instantly. It is clear from the works of \citet{KaibQuinn2008} and \citet{brasser2013} that the emplacement of comets from the planetary region into the Oort Cloud took several hundred Myr. Thus, what we call a primordial Oort Cloud should have been enriched in new members for a time comparable to the cluster lifetime or the full length of our simulations.

Connected to this is another assumption, namely, that the Sun still resides in its birth cluster at the starting time that we use, 100\unit{Myr} after the formation of the Sun and the cluster. For the HM cluster, the risk of ejection of the Sun during this early interval is negligible, but not so for the IM cluster. This has an initial mass of 820\unit{$\msol$}, and 
at the starting time the mass has decreased to 710\unit{$\msol$}. This decrease amounts to 13\%, which we take as a rough estimate of the risk of early ejection of the Sun.

Hence, in the IM cluster case, there is a 13\% chance that some of the comets transferred into the Oort Cloud during the first 100\unit{Myr} as well as all those transferred at later times would not feel any effects of the birth cluster. However, the comets in question -- likely being the majority of all Oort Cloud comets -- would then be emplaced without the help of the birth cluster. The Oort Cloud would then likely have much less of an inner core than if the Sun had stayed in the cluster, and the creation of Sedna-type objects might be strongly curtailed.

With the complementary probability of 87\%, we have underestimated the cluster influence on the early emplaced comets and yet strongly overestimated the influence on those that were emplaced at later times. It seems clear that the overestimates dominate as an error source. Again, one would have to distinguish between two parts of the Oort Cloud -- the comets that were emplaced inside the birth cluster, which did experience its destructive effects, and those that were emplaced after the Sun had left, which did not benefit from the cluster in populating the inner core.

At any rate, we conclude that the Sun could not spend a longer time than approximately 50\unit{Myr} in an IM cluster with an already formed Oort Cloud left intact except under special circumstances. Such circumstances would include a solar orbit which kept the Sun and the Oort Cloud constantly outside the central part of the cluster.

Finally, we stress that there are three categories of birth clusters, which we have estimated to be about equally likely for the Sun in terms of the initial cluster mass function. The LM case is, however, less probable in view of the nucleosynthetic evidence for an early supernova in the neighbourhood (see the discussion above). In the HM case our model should be reasonably good; the IM case was just discussed; and in the LM case the destructive effects of the birth cluster would likely be much smaller.

In fact, we have made a few additional approximations that likely caused us to underestimate the losses of comets, 
thus yielding conservative estimates. 
The first is that we neglected mass segregation in our model cluster. Hence we downplayed the risk for the Sun to encounter a massive star in the high-density environment near pericentre, which would have had dire consequences for the entire Oort Cloud.

The second is that we neglected binary stars. M67 is known to be rich in binaries \citep{richer1998}, and thus other similar clusters may be suspected to be similarly binary-rich. Binary stellar systems are also known from dynamical simulations to form and dissolve within star clusters including clusters of much smaller masses \citep{giersz_statistics_1997}. 
By ignoring binaries, we have artificially increased the number of potential encounter partners of the low mass type, while entirely neglecting a kind of partner that would have had a great capability to transfer energy and momentum to the Sun's motion -- and, similarly, to destroy the Oort Cloud. In particular, ``hard'' binaries in close orbits are known to statistically give energy away to the encounter partner, thus providing an energy source for the dynamical evolution of the cluster, as reviewed by \citet{elson1987}.

A third approximation is that we treat only a rather small number of subsequent encounters in each simulation. This number might be increased without introducing overlapping encounters, but a more fundamental issue is that we select the strongest encounters in terms of the impulse imparted to the Sun per unit mass within the approximation of the classical impulse approximation. 
Certainly, all kinds of encounters may occur, but we systematically disfavour the weaker ones involving the less massive partners. As already remarked, this creates an exaggerated trend for the Sun to move outward in the cluster. Hence, statistically, the Sun spends too much time in the outer, less populated regions, so we underestimate the risk of strong perturbations that characterizes the inner parts. Thus, our treatment should underestimate
the total number of lost comets from the cloud in that a too small number of stellar interactions, and in fact not necessarily the most effectively destructive ones, are considered. Our estimated destruction rates for the Oort Cloud are thus conservative in this respect.

Yet another neglected phenomenon could have increased the number of remaining comets. Some of the lost comets may be picked up by the encountering stars, and this number would likely increase, if we would treat overlapping encounters. Even though most lost comets rather become cluster vagabonds, in due time these could also be picked up by some cluster star, including the Sun, in a way similar to the formation of stellar binaries in clusters.

\subsection{Extra-solar Oort Clouds?}

All this begs the question, whether by assuming that the Sun had a primordial Oort Cloud one should assume that other cluster members of similar types were also equipped with such, primordial cometary clouds. If so, the sort of cluster environment that we consider here might stimulate a certain exchange of comets between different stars \citep{zheng1990,levison2010}. We cannot say how efficient this process would be using our simplified model. On the other hand, it is clear that picked-up Oort Cloud members would be particularly vulnerable to being lost during following encounters, since these would occupy relatively loosely bound orbits.

Concerning the question, if Oort Clouds may be a characteristic feature of Galactic disk stars, we note that \citet{stern_iras_1991} searched for extra-solar Oort clouds around 17 nearby stars by looking for IR excess radiation using IRAS low-resolution data and an S/N-enhancing method, with negative results which, however, may be ascribed to the limited sensitivity. \citet{black_revisiting_2010} has extended this search on the basis of the IRAS sky survey to all F and G dwarfs, augmented with all other stars (then) known to have planetary systems, within 50\unit{pc} distance from the Sun. While challenging, since the sensitivity of the IRAS is again a severely limiting factor, no positive identifications of Oort clouds around other stars were reported.

Some cold, dusty outer disks have been found and studied by means of the Herschel telescope, around young stars, also with planetary systems. One example is the A5~V star HR~8799 with four known planets, at a distance of 40\unit{pc} and an age estimated at 20 to 50\unit{Myr}, which has a central warm dust component, an outer cold component extending from 90 to 300\unit{au} and an outer component of small grains extending beyond 1000\unit{au} \citep{matthews_resolved_2014}. The evidence for any clumping in this halo is, however, meagre. Most observations of debris disks around young stars are limited to A-type stars and to rather small radial distances from the star. One interesting example is, however, the F5/6-type star HD~181327, a member of the 20\unit{Myr} old $\beta$ Pictoris moving group, for which ALMA observations disclose a ring-like CO gas disk, in addition to the dust ring, with a halo extending out to 200\unit{au}, and a CO $+$ CO$_2$ cometary composition \citep{marino_exocometary_2016}.  

Neutron stars capturing comets when passing through the Oort clouds of other stars have been suggested to provide indications on the (non-)existence of extra-solar cometary clouds. \citet{shull_gamma-ray_1995} thus proposed that such events should generate repeating bursts of soft gamma-rays, and estimated that the absence of such Galactic events indicated that at the most a few percent of the Galactic stars have Oort clouds. It is, however, very questionable whether accretion of comets onto neutron stars would occur abruptly enough to generate such bursts; weaker emission, more extended in time and at lower energies, seems more probable (we thank Dr. J. Poutanen for making this point). We conclude that presently no observational limits may as yet be set on how frequent Oort clouds are around stars.

\subsection{Concluding remarks} \label{sec:conclu}

Our basic result is that, if the Sun was born as a member of a relatively rich stellar cluster, a significant part of a primordial Oort Cloud would not likely survive the time in the cluster until the Late Heavy Bombardment. The relative extent of this depletion depends on the detailed orbital evolution of the Sun within the cluster. In the case of a massive depletion, the formation of the present Oort Cloud as a consequence of a late planetary migration within the Nice Model \citep{Levison2011} appears to be a viable scenario, provided that the Sun had then already left the cluster, or was about to do so. Such a scenario has been explored by \citet{brasser2013}.

In a very massive cluster the escape of primordial Oort Cloud comets is mainly caused by the disrupting effect of stellar encounters. The eccentricity pumping due to the cluster tide plays a role only for a minority of comets in the inner part of the cloud. In the outer part the energy perturbations caused by the tide may constitute an important source of comet losses. In clusters of lower mass the latter type of tidal perturbations provide a major loss mechanism in all parts of the Oort Cloud. If the cluster mass is very high, the number of comets penetrating to within 5\unit{au} of the Sun amounts to $\simeq 4$\unit{\%} of the initial, inner cloud members. For the lower cluster mass, this fraction decreases to about 2\unit{\%}. In case the primordial Oort Cloud had a high mass, this might cause an important cometary bombardment of the terrestrial planets (including a late veneer of H$_2$O) well before the time of the LHB.

The influence of the stellar encounters on the solar orbit can be seen in Fig.~\ref{fig:sorb}. From this and the solar escape rate, we conclude that the Sun typically receives a cumulative impulse of several km/s. Since the number of encounters per simulation is only about 20, in a random walk there must be some individual kicks experienced by the Sun that amount to about 1\unit{km/s} and are thus strong enough to unlink most of the Oort Cloud comets. The solar impact parameters of those stars are typically much less than the Sun-comet distances. Thus, the main mechanism of comet escape in our model is that the Sun is kicked away from its comet cloud rather than individual comets being kicked away from the Sun.

By restricting our simulations to a small number of stellar encounters, we may have introduced too large a statistical dispersion of the results. In particular, the simulations that left the Oort Cloud -- particularly for the intermediate-mass cluster -- with much more comets than the median can be regarded as chance selections of solar orbits with weaker tidal effects and encounter sequences than normal. We have shown that the minimum periapsis distance $q_{\rm min}$ of the Sun plays a decisive role for the survival frequency in the Oort Cloud -- the larger $q_{\rm min}$, the more survivors. According to our results, the category of outcomes with the larger survival rate makes a significant though not dominant contribution, but some caution is warranted, especially concerning the results for the IM case, until more realistic cluster simulations can be made. Tentatively, there is no reason to suspect that we would have exaggerated the comet loss rate in the IM cluster by selecting initial solar orbits with too small $q_{\rm min}$. The selection was made on the basis of the general density profile of the cluster, so we neglected the fact that the Sun, being a relatively high-mass star, would tend to prefer the central region at the time we started our simulations and the cluster was already considerably relaxed. Again, however, only fully realistic simulations including binarity and other special phenomena will provide the full answer to this question.

A primordial Oort Cloud may consist of comets that originated in the accretion zone of the giant planets. However, if the planetary orbital instability of the Nice Model actually happened very early, as was recently argued by \citet{kaib_fragility_2016}, the Oort Cloud originating from the trans-planetary disk would also be primordial. It may have been formed after the birth cluster was dispersed, if this cluster contained only about a hundred stars. Such a birth cluster may not be totally excluded by the argument of nucleosynthesis providing the short-lived radionuclides for the solar nebula, even though somewhat special circumstances would be called for in the case of the Sun. If so, the formation scenario of \citet{brasser2013} would be a relevant model except that the timing of the event is different.

On the other hand, if the birth cluster was of the IM or HM kind, we have shown that the primordial Oort Cloud would largely survive only as a tight, inner core. Models of Oort Cloud formation in such a dense stellar environment indeed predict an initial cloud structure dominated by such a core \citep{brasser2006}. Therefore, in such a case the existence of the present outer halo, from where the observed comets can be transferred by the Galactic tide, requires a mechanism of energy transfer that can activate the core from its inert state. The alternative would be a late planetary instability as investigated by \citet{brasser2013}, in which case it is reasonable to assume that the Sun had already left the birth cluster.

To be specific, taking the Galactic disk tide to be the mechanism for bringing Oort Cloud comets into the inner solar system, the tidal torquing time scale \citep{heisler1986} is found to be longer than the age of the solar system, unless $a\gtrsim 5\,000$\unit{au}. Stellar perturbations would not change this result drastically. We find that such comets in general would not survive the dwelling time within the birth cluster. A fossilized, primordial inner core of comets with $a \lesssim 3\,000$\unit{au} might exist, possibly including the Sedna-type objects, but it would not be able to produce the observed, new Oort Cloud comets. The gap in orbital energy would have to be bridged by an as yet unidentified mechanism.

Returning to the issue of extra-solar Oort Clouds, our results suggest that such clouds would often be severely truncated by the effects of the birth clusters and therefore much smaller than usually thought. Quite likely, Oort Clouds in general would mostly stem from trans-planetary planetesimal disks, and thus, their existence depends on the way extra-solar planets have migrated.



\begin{acknowledgements}
This work was supported by the Polish National Science Center under Grant No. 2011/01/B/ST9/05442 as well as Grant No. 74/10:2 of the Swedish National Space Board. 
T.N. acknowledges support from Swedish National Space Board (Rymdstyrelsen) and Anna \& Allan L\"ofbergs stiftelse.
Our thanks go to Ross Church, Mirek Giersz, Alessandro Morbidelli and Juri Poutanen for enlightening discussions, and to the referee, Nathan Kaib, for raising points of crucial importance for the interpretation of our results.
Computations were performed on resources provided by SNIC through Uppsala Multidisciplinary Center for Advanced Computational Science (UPPMAX), Lund University Center for scientific and technical computing (LUNARC), and High Performance Computing Center North (HPC2N), under projects s02011-10, snic2013-11-24, and SNIC2015-1-309.
\end{acknowledgements}

\bibliography{references}



\appendix

\section{Calculation of stellar cluster structure} \label{sec:stellarstructure}

\begin{table*} \centering
\caption{Input parameters and resulting properties of the stellar clusters. The properties are represented in terms of the input parameters $a$, $b$ and $U_0$ (see Sect.~\ref{sec:cl-prop}).} \label{tbl:modelparams}
\begin{tabular}{c|ccc|rrccrr} \hline\hline \noalign{\smallskip} 
& \multicolumn 3 {c|} {Input parameters} & \multicolumn 6c {Model properties} \\
Time span & $a \times 10^{31} $ & $b \times 10^{7}$ & $U_0 \times 10^{-5}$ & \multicolumn1c{$\mass_\text{cl}$} & \multicolumn1c{$R_{\rm t}$} & \multicolumn1c{$r_\text{h}$} & \multicolumn1c{$R_\text{c}$} & \multicolumn1c{$\mean {\rho_{\rm h}}$} & \multicolumn1c{$\mean {\rho_{\rm c}}$} \\
($\rm Myr$) & ($\rm s^3\, kg^{1/2}\, m^{-6}$) & ($\rm J^{-1}$) & ($\rm m^2 \, s^{-2}$) & \multicolumn1c{($\msol$)} & \multicolumn1c{($\rm pc$)} & ($\rm pc$) & ($\rm pc$) & \multicolumn1c{($\msol\,\text{pc}^{-3}$)} & \multicolumn1c{($\msol\,\text{pc}^{-3}$)} \\
\hline \noalign{\smallskip} 
\multicolumn{10}c{The high-mass (HM) cluster} \\ 
\hline 
\noalign{\smallskip} 
100--500 &  $4.036$ & $3.632$& $-162.3$& 13821& 28.8&  4.87&  1.88&  14.3&    57.6 \\
\hline \noalign{\smallskip} 
\multicolumn{10}c{The intermediate-mass (IM) cluster} \\ 
\hline 
\noalign{\smallskip} 
100--125 &  $193.4$&  $11.15$& $-75.83$&   692& 13.0&  1.88&  0.07& 119.5& 11171.7 \\
125--150 &  $228.8$&  $11.98$& $-76.24$&   632& 12.7&  1.00&  0.04&  75.0& 24678.8 \\ 
150--175 &  $203.4$&  $14.59$& $-63.03$&   580& 12.3&  1.13&  0.05&  48.1& 16269.1 \\ 
175--200 &  $181.6$&  $17.44$& $-52.52$&   535& 12.0&  1.23&  0.05&  34.2& 10399.9 \\
200--225 &  $162.5$&  $20.70$& $-43.71$&   492& 11.7&  1.31&  0.06&  26.1&  6527.4 \\ 
225--250 &  $151.8$&  $23.92$& $-37.49$&   453& 11.3&  1.37&  0.07&  21.0&  4490.9 \\ 
250--275 &  $144.0$&  $27.09$& $-32.71$&   420& 11.0&  1.40&  0.07&  18.2&  3193.8 \\ 
275--300 &  $135.3$&  $31.19$& $-27.97$&   385& 10.7&  1.43&  0.08&  15.6&  2091.7 \\ 
300--325 &  $151.6$&  $32.59$& $-27.13$&   355& 10.4&  1.42&  0.09&  14.9&  2243.6 \\ 
325--350 &  $155.1$&  $37.97$& $-23.19$&   312& 10.0&  1.43&  0.09&  12.7&  1800.4 \\ 
350--375 &  $157.2$&  $42.21$& $-20.72$&   285&  9.7&  1.43&  0.10&  11.6&  1390.0 \\ 
375--400 &  $154.4$&  $48.34$& $-17.65$&   256&  9.3&  1.40&  0.09&  11.1&   991.8 \\ 
400--425 &  $171.7$&  $54.25$& $-15.97$&   224&  9.0&  1.42&  0.10&   9.4&   947.6 \\ 
425--450 &  $174.1$&  $64.02$& $-12.92$&   194&  8.6&  1.32&  0.10&  10.0&   567.2 \\
450--475 &  $192.8$&  $68.46$& $-11.78$&   175&  8.2&  1.24&  0.13&  11.1&   434.4 \\
475--500 &  $201.1$&  $84.04$& $-9.141$&   149&  7.8&  1.22&  0.15&   9.8&   223.5 \\ 
500--525 &  $238.1$&  $94.20$& $-7.963$&   127&  7.3&  1.15&  0.16&  10.1&   185.6 \\ 
\hline \noalign{\smallskip}
\end{tabular}
\end{table*}

We prescribe that the model cluster should have a given mass $\mass_{\rm cl}$ and be situated in the Galactic potential at distance $r_G$ from the Galactic centre. This allows the calculation of a tidal radius of the cluster, $R_t$. Using a limiting energy $E_t$, taken to be the energy of a circular orbit situated at a distance from the cluster centre $r = R_t$, 
our \citeauthor{king1966} model is calculated from a distribution function $\varphi$, 
which can be expressed as 
\begin{align}
	\varphi(E) &= 
		\begin{cases}
			a \left\{e^{b(\tidalE - \totE)} - 1 \right\} & \totE < \tidalE \\
			0  & \totE \ge \tidalE \\
		\end{cases} 
\label{eq:densityfunction}
\end{align}
where $\totE = U + v^2/2$ is the total energy of a test particle, and $a$ and $b$ are positive constants of the model.
We rewrite this distribution function when $\totE < \tidalE$ as a function of velocity and radial position, 
\begin{align}
\varphi(v, r) &= a \left\{e^{\frac b 2 \vmax(r)^2} e^{-\frac b 2 v^2} -1\right\} \label{eq:king1}
\end{align}
and solve for the density $\rho$,
\begin{align}
	\rho(r) 
	 & = 4\pi \int_0^{\vmax(r)} \varphi(v,r) v^2 dv \nonumber\\
	 &= 4 \pi a \int_0^{\vmax(r)} v^2 \left\{e^{\frac b 2 \vmax[^2](r)} e^{-\frac b 2 v^2} -1\right\} dv  \nonumber\\
	 &= 4 \pi a \left\{ e^{\frac b 2 \vmax[^2](r)} \int_0^{\vmax(r)} v^2 e^{-\frac b 2 v^2} dv - \frac{{\vmax[^3](r)}}3 \right\}. \label{eq:king2}
\end{align}
The velocity distribution can be evaluated by multiplying \eqref{eq:king1} by $4 \pi v^2$, giving locally a truncated Maxwellian distribution.
The structure of the cluster, its distribution of mass and the gravitational potential, is finally computed by solving the integral \eqref{eq:king2} simultaneously with Poisson's equation, 
\begin{align}
	\frac{d^2 U}{d r^2} + \frac 2 r \frac{d U}{d r} &= 4 \pi G \rho
\end{align}
using the parameter substitutions suggested by \citet[eqns. 19--22]{king1966}. 
The equations are solved from the cluster centre, where the potential energy $U_0$ is taken as a free parameter.
The free parameters $a, b, U_0$ thus represent a reformulation of the classical free parameters $W_0$, $r_0$ and $\mass_{\rm cl}$ required to fit a \citeauthor{king1966} model to an idealized stellar cluster.

The input parameters and resulting properties of the optimal model are given in Table~\ref{tbl:modelparams}.

\section{Implementation of stellar encounters} \label{sec:stellarencounters}

Each list covers a time interval of 40\unit{Myr}, and it comes from a preliminary integration of the solar orbit in steps of 3\unit{kyr}, yielding the mean encounter frequency, $\langle n \langle v_{\rm rel}\rangle\rangle$, where $n$ and $\langle v_{\rm rel}\rangle$ are interpolated from the solution shown in Fig.~\ref{fig:encounter_frequency}. 
For each step, we let an encounter occur within $b_{\rm max} = 20\,000$\unit{au}, if
\begin{align}
   b_{\rm max}^2 \pi  \int_{t_i}^{t_{i+1}} n \langle v_{\rm rel} \rangle dt > \xi
\end{align}
where $t_{i+1} = t_i + 3$\unit{kyr}, and $\xi \in [0,1]$ is a random number drawn for each step from a uniform distribution. This typically generates about 100 encounters. For each such encounter, we generate a random stellar mass from the IMF as described in Sect.~\ref{sec:cl-stars}, evolving it to the age of the cluster at the current time. If the star is found to have evolved through a supernova phase, it is discarded and a new value is drawn from the IMF and again evolved to the current time. The impact parameter for each selected encounter is determined randomly using $b = b_{\rm max} \sqrt \xi$, for a new random $\xi \in [0,1]$. Finally, the star is assigned a velocity $v_{\rm rel}$ relative to the Sun as in Sect.~\ref{sec:encounterflux}.

After this preliminary selection, we know the times and parameters of all the potential, upcoming stellar encounters, and we are able to pick one of them based on the $S$ values. When modelling the encounter in a two-body scattering problem, the approximation would be to let the encountering star aim from infinity at a position on a heliocentric circle with radius $b$ in a plane perpendicular to the direction of approach (called the impact plane). However, under the influence of the cluster potential, this straight-line approximation cannot be used. 
Instead, we realise the closest approach by first choosing randomly the direction of relative motion, which defines the impact plane, and then placing the star on the impact plane at a distance $b$ from the Sun at a random azimuth.

The orbits of both stars are then integrated backward in time in the cluster potential with no mutual gravitational interaction, until the distance between the stars is $d_\text{start}$. The time and geometric configuration at this stage are stored as the initial state of that encounter. In the few cases of very slow encounters, where the backward integration overlaps the previous encounter, the setup is considered as failed and is discarded. 
We then select the second largest value of $S$ and repeat the calculation of the initial state.

\end{document}